\documentclass[12pt]{article}

\usepackage[margin=0.85in]{geometry}
\usepackage[utf8]{inputenc}
\usepackage[T1]{fontenc}
\usepackage{textcomp}
\DeclareUnicodeCharacter{2212}{$-$}
\usepackage{booktabs}
\usepackage{dcolumn}
\usepackage{array}
\usepackage{longtable}
\usepackage{caption}
\usepackage{subcaption}
\usepackage{setspace}
\usepackage{amsmath}
\usepackage{amssymb}
\usepackage{hyperref}
\usepackage{natbib}
\usepackage{xcolor}
\usepackage{microtype}
\usepackage{enumitem}
\usepackage{tabularx}
\usepackage{rotating}
\usepackage{pdflscape}
\usepackage{multirow}
\usepackage{threeparttable}
\usepackage{siunitx}
\usepackage{graphicx, amsfonts}

\hypersetup{
    colorlinks=true,
    linkcolor=blue!60!black,
    citecolor=blue!60!black,
    urlcolor=blue!60!black
}

\newcolumntype{d}[1]{D{.}{.}{#1}}

\newcommand{\appendixsection}{
    \setcounter{section}{0}
    \setcounter{table}{0}
    \setcounter{figure}{0}
    \renewcommand{\thesection}{A.\arabic{section}}
    \renewcommand{\thetable}{A.\arabic{table}}
    \renewcommand{\thefigure}{A.\arabic{figure}}
}

\begin{document}

\title{Degrees of Devaluation: College Expansion and the Credential Trap in India}

\author{Kaibalyapati Mishra\thanks{Centre for Economic and Social Studies (CESP), Institute for Social and Economic Change (ISEC), Bengaluru, India. E-mail: \texttt{kaibalyapati@isec.ac.in}. All errors are my own.}\\ Centre for Economic and Social Studies (CESP)\\ Institute for Social and Economic Change (ISEC), Bengaluru, India}

\date{\today}

\maketitle

\begin{abstract}
\noindent India's post-liberalisation higher education expansion was premised on widening credential access for historically excluded groups. We show that the groups most expected to benefit --- Scheduled Caste and Scheduled Tribe (SC/ST) workers --- instead bore a disproportionate share of the resulting wage cost, a pattern we term the \textit{double whammy}. We merge eight rounds of the NSS Employment--Unemployment Survey (1987--2011) with a district-level measure of college-expansion intensity built from the All India Survey on Higher Education (AISHE) and estimate reduced-form triple- and quadruple-difference wage specifications across 91 districts in six states ($N = 79{,}904$), interacting graduate status, expansion intensity, and post-expansion cohort. The human capital return to a degree remains large and positive throughout (about 1.08 log points), yet the graduate wage premium erodes for post-2004 cohorts in high-expansion districts: non-SC/ST graduates earn roughly 9 per cent less than comparable graduates in low-expansion districts at mean intensity, and SC/ST graduates face an additional penalty of about 34 per cent (a combined shortfall near 43 per cent). The SC/ST differential is statistically indistinguishable from zero before the expansion and emerges only afterwards. Non-graduate placebo and pre-trend tests are broadly consistent with a credential-signalling channel, though we flag the limits of the design rather than claim clean identification. The results suggest that expanding access without commensurate investment in institutional quality can deepen, rather than narrow, labour-market inequality for disadvantaged groups.\\[0.8em]
\end{abstract}
\onehalfspacing

\pagebreak
\section{Introduction}
\label{sec:introduction}

Indian higher education was, for most of the post-independence period, the preserve of a narrow elite \citep{kumar1985reproduction}. Economic liberalisation, and the deregulation of private providers that followed the Supreme Court's ruling in \textit{T.M.A. Pai Foundation v. State of Karnataka} \citep{tmapai2002}, changed this within a generation: a small, overwhelmingly public system became a large and predominantly private one, and a credential once confined to a narrow stratum was extended, in principle, to social groups long excluded from it (Section~\ref{sec:background}). This expansion was framed almost without exception as democratising. This paper asks whether that promise was kept, and for whom, and finds an answer more troubling than the framing suggests.

The promise was never self-evidently kept, because expanding the supply of a credential sets two forces against each other. More colleges mean more schooling and a larger human capital stock, which should raise productivity and wages. But a credential is valuable to employers partly because it is informative, conveying something about an applicant's unobservable ability, and that informativeness depends on scarcity. When graduates multiply faster than employers can distinguish strong credentials from weak ones --- and more so when much of the new supply comes from institutions of uncertain and uneven quality --- the screening value of the degree falls, and the wage premium it commands falls with it. Expansion can therefore raise the average worker's human capital and erode the informational return to the very credential it makes more accessible, at the same time.

This erosion is not borne evenly, and that unevenness is the central concern of the paper. The loss of signalling value falls most heavily on graduates whose credentials already face the greatest employer scepticism, and Scheduled Caste (SC) and Scheduled Tribe (ST) workers are the clearest case in the Indian context: lacking the networks, social capital, and inherited occupational standing available to higher-caste workers, they have relied on the formal credential as their primary, often sole, means of signalling ability in the labour market. Statistical discrimination theory \citep{phelps1972, arrow1973b} gives this a sharp prediction. As the average credential becomes less informative, employers rely more heavily on group-level priors, and the discount applied to workers presumed more likely to hold a lower-quality post-expansion degree grows sharpest for those who had the least to fall back on. The credential intended as the great equaliser becomes, under these conditions, a trap --- a pattern we term the \textit{double whammy}.

We test this using eight rounds of the NSS Employment--Unemployment Survey (1987--2011), merged with a district-level measure of expansion intensity constructed from twelve rounds of the All India Survey on Higher Education (AISHE, 2010--11 to 2021--22). For each of 91 districts across six states, the measure records the stock of colleges established after 2004 relative to the pre-1995 baseline, isolating the large private-led expansion wave that followed deregulation. We interact this intensity with individual graduate status, post-expansion cohort, and SC/ST status to test the double whammy directly. We state plainly at the outset that this is a reduced-form design: an earlier attempt to use expansion intensity as an instrument for individual graduate status failed the relevance condition, and we therefore treat the expansion measure as a continuous moderator of the graduate wage premium rather than as a source of exogenous variation in who becomes a graduate. Section~\ref{sec:empirical} and the Online Appendix set out this choice and its limits in full.

The findings are consistent with this framework and, in their distributional dimension, substantial. The human capital return to a degree remains large and positive throughout the period (about 1.08 log points in the full graduate--non-graduate gap), consistent with expansion raising the educated workforce's productivity. The graduate wage premium nonetheless erodes for post-2004 cohorts in high-expansion districts: at mean expansion intensity, non-SC/ST graduates in these cohorts earn roughly 9 per cent less than comparable graduates in low-expansion districts. SC/ST graduates face an additional penalty of about 34 per cent on top of this, for a combined shortfall near 43 per cent. The SC/ST differential is statistically indistinguishable from zero before the expansion and emerges only afterwards, a pattern we read as consistent with the signalling mechanism without overstating it: our placebo and pre-trend tests broadly support a credential-based interpretation, though, as we report transparently, not uniformly cleanly.

That degree returns have fallen in India as the graduate pool has grown is by now established \citep{azam2016, kumar2008}, including in an earlier and shorter treatment of the aggregate question \citep{mishra2024}.\footnote{\citet{mishra2024} documents aggregate degree devaluation in India using a cohort-based design with a single 1995 cutoff. This paper differs in tying devaluation to a measurable district-level expansion gradient rather than one national break, and in making the \textit{distributional} incidence of that devaluation across social groups --- which the earlier work does not address --- its central object.} What has not been shown is that the cost of devaluation falls disproportionately on the historically disadvantaged, or that the SC/ST graduate wage gap is itself a product of the expansion rather than a pre-existing feature of the labour market. By locating both the level and the distribution of devaluation along a single supply-shock gradient, and by showing that the caste differential is absent before the expansion and substantial after it, the paper recasts credential devaluation as a question of distribution rather than of average returns alone --- one with direct bearing on how access-widening education policy interacts with labour-market discrimination.

The paper proceeds as follows. Section~\ref{sec:background} describes the institutional context of India's higher education expansion. Section~\ref{sec:literature} situates the analysis within the literatures on returns to education, credential inflation, and statistical discrimination. Section~\ref{sec:theory} sets out a simple framework linking expansion to the differential erosion of the signal. Section~\ref{sec:data} describes the data and sample construction. Section~\ref{sec:empirical} details the empirical strategy, identification, and its limits. Section~\ref{sec:results} presents the main results, robustness and placebo tests, and heterogeneity analysis. Section~\ref{sec:discussion} discusses mechanisms and limitations. Section~\ref{sec:conclusion} concludes with policy implications.
\section{Institutional Background}
\label{sec:background}

India's higher education system was, until the early 1990s, overwhelmingly public, tightly regulated, and small relative to the size of the country's youth population. Colleges were established and funded predominantly by state governments and university grants, fees were nominal and centrally constrained, and admission was rationed by a combination of scarce capacity and academic selection. The system reproduced existing social hierarchies more than it disturbed them: access was concentrated among urban, upper-caste, and propertied households, and the degree functioned as a relatively reliable marker of a narrow and advantaged stratum \citep{kumar1985reproduction}. Caste-based reservation in public institutions, in place since independence, widened access at the margin for Scheduled Caste and Scheduled Tribe students, but the absolute scale of the system remained too small for this to transform the composition of the graduate workforce.

The structural change came with liberalisation and, decisively, with the judicial reinterpretation of the rights of private educational providers. The Supreme Court's ruling in \textit{T.M.A. Pai Foundation v. State of Karnataka} \citep{tmapai2002} affirmed that private unaided institutions had a substantial right to administer themselves, including latitude over fees and admissions, subject only to limited regulatory oversight. Together with subsequent rulings and a permissive stance from the University Grants Commission, this established the legal and economic conditions for a private-college boom. The number of degree-granting colleges, broadly stable through the 1980s, rose from roughly 7,500 in 1990 to over 40,000 by 2015, accelerating through the late 1990s and then expanding sharply from the mid-2000s, with the bulk of the new supply taking the form of private self-financing institutions charging market-determined fees. By the second decade of the 2000s the majority of colleges were private and unaided, a structural inversion of the system that had prevailed a generation earlier.

This expansion was, in principle, democratising. It dramatically loosened the capacity constraint that had rationed access, extended the geographic reach of higher education into districts that had never had a local college, and opened the credential to first-generation entrants from social groups long excluded from it. The policy literature and public discourse alike framed the private-college boom as an engine of inclusion, and the affirmative-action literature reinforced this reading: studies of reservation in selective public institutions found that widening access for disadvantaged groups raised their earnings without imposing measurable costs on others \citep{bertrand2010}.

\begin{figure}[htbp]
\centering
\includegraphics[width=0.82\textwidth]{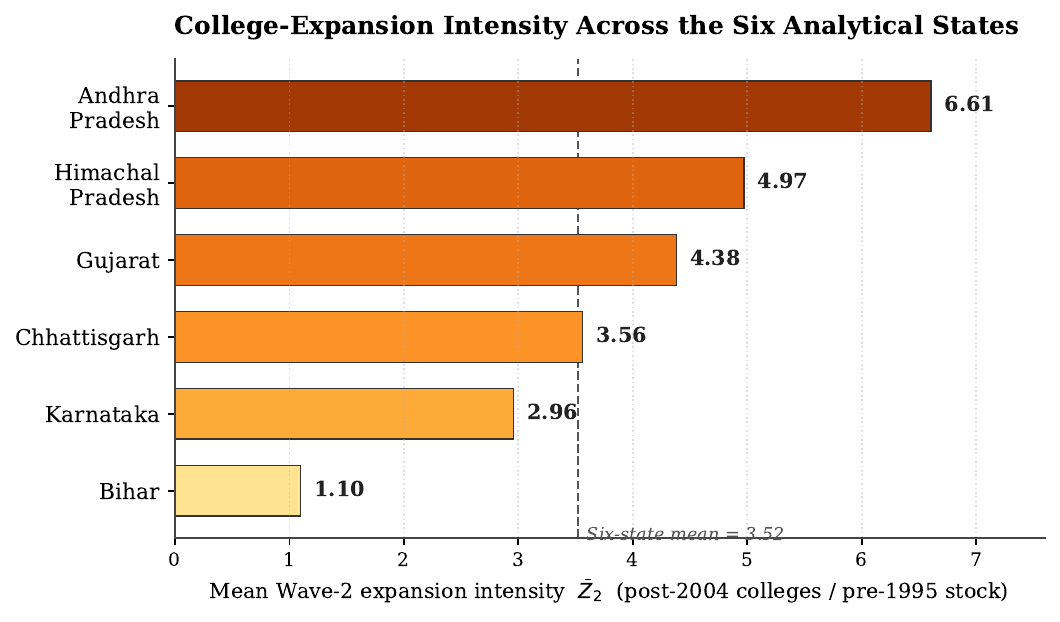}
\caption{College-Expansion Intensity Across the Six Analytical States}
\label{fig:z2_by_state}
\begin{minipage}{0.82\textwidth}
\vspace{0.4em}
\footnotesize \textit{Notes:} Mean Wave-2 expansion intensity $\bar{Z}_2$ by state, defined as the district-level ratio of colleges established after 2004 to the pre-1995 baseline college stock, averaged across individuals in the analysis sample ($N = 79{,}904$). The dashed line marks the six-state mean ($\bar{Z}_2 = 3.52$). State means are respondent-weighted and correspond to Panel D of Table~\ref{tab:summary_stats}; unweighted district-level means are reported in the Online Appendix. Bihar's low and weakly varying intensity is the basis for its exclusion from the state-level heterogeneity analysis in Section~\ref{sec:results}.
\end{minipage}
\end{figure}

The expansion was also, however, highly uneven in both its intensity and its institutional character, and it is this unevenness that our analysis exploits. The new private colleges varied enormously in quality, from a small number of well-resourced institutions to a long tail of thinly capitalised establishments whose degrees carried little assurance of underlying skill. The pace of expansion differed sharply across states and districts: some regions saw the local stock of colleges multiply several times over within a decade, while others, constrained by weaker household demand and thinner regulatory capacity, saw comparatively modest growth. Figure~\ref{fig:z2_by_state} illustrates this cross-state heterogeneity in our six analytical states, measured as the ratio of colleges established after 2004 to the pre-1995 baseline stock. Andhra Pradesh, with a long history of private professional education, experienced the most intense expansion, with post-2004 establishments numbering more than six times the pre-1995 stock on average; Bihar, by contrast, saw the local college base grow by only a fraction of that, reflecting weaker effective demand and a less developed private sector. Gujarat, Himachal Pradesh, Chhattisgarh, and Karnataka fall between these extremes. This cross-sectional and temporal variation in expansion intensity is the variation our empirical strategy exploits, and we describe its construction in Section~\ref{sec:data}.

\section{Literature Review}
\label{sec:literature}

This paper speaks to four bodies of literature: the returns to education in developing countries, the economics of credential inflation, statistical discrimination in labour markets, and the political economy of access-widening education policy in India.

\subsection{Returns to Education and Credential Inflation}

The canonical Mincerian framework decomposes the wage return to education into a human capital component --- the productivity enhancement from skills acquired in school --- and a signalling component --- the informational value of the credential to employers \citep{mincer1974, spence1973}. A large literature documents positive returns to schooling across developing countries, with India showing returns of approximately 7--10 per cent per year of schooling in the 1990s \citep{duflo2001, psacharopoulos2004}. More recent work, however, documents declining wage returns to education in India as the educated workforce has expanded \citep{kumar2008}, consistent with a credential devaluation hypothesis.

The theoretical mechanism linking educational expansion to credential devaluation was formalised by \citet{arrow1973} and \citet{spence1973}: when education primarily serves as a signal of pre-existing ability rather than as a generator of productive human capital, an expansion in the supply of credentials dilutes the informational content of the signal. \citet{dore1976} coined the term ``diploma disease'' to describe this dynamic in developing-country contexts more broadly, where credentials proliferate faster than the labour market's capacity to use them productively, and \citet{caplan2018} provides a comprehensive modern treatment of the same argument. A separate empirical literature on sheepskin effects, beginning with \citet{hungerford1987}, isolates the discrete jump in wages at credential thresholds as evidence that signalling, not just accumulated skill, is being rewarded --- a distinction we draw on directly in separating the human capital and signalling channels in Section~\ref{sec:theory}. \citet{bedard2001} offers empirical evidence on the labour-market consequences of US community college expansions specifically. The closest antecedents to our analysis in the Indian context are \citet{mishra2024}, who establishes the existence of degree devaluation in India using a cohort-based design with a 1995 expansion cutoff, and \citet{azam2016}, who documents falling returns to graduate education in Indian household data.

\subsection{Higher Education Expansion and Labour Market Outcomes}

The supply-side literature on higher education expansions and labour market outcomes identifies two competing mechanisms. \citet{card1995} and \citet{duflo2001} use geographic variation in institutional access as instruments for individual schooling, finding positive wage returns consistent with human capital theory. More recent work exploiting rapid expansion episodes documents heterogeneous effects: \citet{carneiro2011} find that marginal returns to college in the US are positive but declining, while \citet{liu2016} document wage compression following China's higher education expansion. For India specifically, \citet{zimmermann2019} finds that access to engineering colleges raises graduate wages in the short run but depresses non-graduate wages through displacement.

\subsection{Statistical Discrimination and Social Identity}

Our double whammy mechanism builds on the statistical discrimination literature initiated by \citet{phelps1972} and \citet{arrow1973b}. In this framework, employers use group membership as a proxy for unobservable individual productivity when signals are noisy. When the signal quality of a credential declines for all groups, the discount applied to credentials from groups perceived as lower-ability is amplified, because the marginal information content of the credential --- already low --- is further reduced for those groups. \citet{coate1993} formalise conditions under which this produces a self-confirming equilibrium where disadvantaged groups invest less in signalling, consistent with the cohort-level decline in graduate shares we document. In the Indian context, \citet{hnatkovska2012} and \citet{thorat2010} document persistent wage gaps for SC/ST workers that cannot be explained by observable human capital differences, consistent with a discrimination channel, while \citet{deshpande2011} provides the foundational account of how the absence of caste networks and inherited occupational capital leaves SC/ST workers disproportionately reliant on formal credentials as a signalling device in the first place --- the premise on which our double whammy mechanism rests.

\subsection{Access-Widening Policy and Reservation in Higher Education}

A separate literature examines the labour-market consequences of policies that widen access to higher education for disadvantaged groups specifically, of which India's caste-based reservation system is the leading example. \citet{bertrand2010} show that reservation in engineering college admissions in India raises earnings for beneficiaries without measurable displacement costs for non-reserved students, a finding generally read as supportive of access-widening policy on its own terms. Our results complicate this picture from a different angle: even where access expands without an explicit quota mechanism, as in the private-college wave we study, the wage benefit of that access can be eroded unevenly across exactly the groups such policies are meant to help, which speaks directly to the policy discussion in Section~\ref{sec:conclusion}.

\section{Theoretical Framework}
\label{sec:theory}

\subsection{Setup}

Consider a labour market in which a worker's ability $\theta_i$ is unobservable to employers and distributed, within social group $g \in \{A, B\}$, as $\theta_i \sim \mathcal{N}(\mu_g, \tau^2)$, where group $B$ comprises SC/ST workers. Workers choose whether to acquire a degree at cost $c_i$, with $c_H < c_L$ in the standard signalling fashion \citep{spence1973, arrow1973}, so that the degree carries information about ability. We depart from the textbook model in one respect that captures the central feature of the Indian setting: beyond the degree, employers also observe a noisy auxiliary signal of ability, $s_i = \theta_i + \varepsilon_i$, where $\varepsilon_i \sim \mathcal{N}(0, \sigma_g^2)$. This auxiliary signal stands for the secondary credentialling channels --- institutional prestige, professional and caste networks, parental occupation --- that allow employers to distinguish strong from weak candidates within the pool of degree-holders. The one substantive asymmetry we impose is that this auxiliary signal is noisier for the disadvantaged group,
\begin{equation}
    \sigma_B > \sigma_A,
    \label{eq:sigma}
\end{equation}
formalising the well-documented fact that SC/ST workers have systematically less access to such networks and inherited occupational capital, and must rely on the formal credential more heavily as a result \citep{deshpande2011, thorat2010}. Everything that follows is a consequence of this single primitive.

\subsection{Wages and Bayesian Updating}

Facing a group-$g$ graduate with auxiliary signal $s$, a competitive employer pays the posterior expected ability. With Gaussian signals this is the standard precision-weighted average of the group prior mean and the auxiliary signal:
\begin{equation}
    w_g(s) = \mathbb{E}[\theta \mid s, g]
           = \lambda_g \, \mu_g + (1 - \lambda_g)\, s,
    \qquad
    \lambda_g = \frac{\sigma_g^2}{\sigma_g^2 + \tau^2}.
    \label{eq:posterior}
\end{equation}
The weight $\lambda_g$ placed on the group prior is increasing in the noisiness of the auxiliary signal, $\partial \lambda_g / \partial \sigma_g^2 > 0$: when an employer can learn little about a worker from the secondary signal, they fall back more heavily on what they believe about the worker's group. Equation~\eqref{eq:sigma} therefore implies
\begin{equation}
    \lambda_B > \lambda_A,
    \label{eq:lambda}
\end{equation}
the disadvantaged group's wage is more sensitive to the group prior and less responsive to individual signals --- the formal counterpart of statistical discrimination operating more strongly where individual information is scarcer \citep{phelps1972, coate1993}.

\subsection{The Expansion Effect}

Higher education expansion in district $d$ at intensity $Z_d$ lowers the perceived average quality of the credential pool --- whether because the marginal entrant is less able, or because newly established institutions deliver lower average instructional quality, or both. We capture this as a decline in the group prior mean that employers attach to graduates, $\partial \mu_g / \partial Z_d < 0$, common across groups.\footnote{What matters for the asymmetry result is only that expansion lowers the prior that employers bring to graduates of both groups; whether it does so through the ability composition of entrants or through institutional quality is immaterial to the comparative static, and our data cannot separate the two. We treat $\mu_g$ as perceived average quality rather than a literal high-ability share for this reason, and return to the distinction in the conclusion.} Differentiating the wage in equation~\eqref{eq:posterior} with respect to $Z_d$ gives
\begin{equation}
    \frac{\partial w_g}{\partial Z_d} = \lambda_g \, \frac{\partial \mu_g}{\partial Z_d} < 0,
    \label{eq:dwdz}
\end{equation}
so wages fall with expansion for both groups. The magnitude of the decline, however, is scaled by $\lambda_g$, and since $\lambda_B > \lambda_A$ by equation~\eqref{eq:lambda}, the same deterioration in perceived credential quality produces a strictly larger wage loss for the disadvantaged group:
\begin{equation}
    \left| \frac{\partial w_B}{\partial Z_d} \right|
    > \left| \frac{\partial w_A}{\partial Z_d} \right|.
    \label{eq:prediction}
\end{equation}
This is the \textit{double whammy}, and it is now a result rather than an assumption: both groups suffer credential devaluation under expansion, but because employers lean more heavily on the group prior precisely where individual signals are noisiest, SC/ST graduates absorb a disproportionate share of the loss. The mechanism requires no assumption that expansion differentially changes who enters the graduate pool by group; it requires only that the disadvantaged group's auxiliary signal is noisier to begin with, equation~\eqref{eq:sigma}, which is the formal content of the network and credentialling disadvantage documented in Section~\ref{sec:literature}. Equation~\eqref{eq:prediction} is the central prediction we take to the data: the coefficient on \texttt{graduate} $\times Z_d \times$ \texttt{post} should be negative (base devaluation), and the additional interaction with \texttt{disadvantaged} should be negative as well (the double whammy).

\subsection{Human Capital vs. Signalling}

The framework is consistent with a positive unconditional return to education (human capital channel) coexisting with a negative interaction between expansion intensity and the graduate wage premium (signalling channel). The two mechanisms operate simultaneously and are identified separately in our empirical specification: the coefficient on \texttt{graduate} captures the average human capital return, while the coefficient on \texttt{graduate $\times$ $Z_d$ $\times$ post} captures the erosion of the signalling premium under expansion. We note, however, that the wage results cannot by themselves separate the compositional and institutional-quality routes through which perceived credential quality $\mu_g$ falls; as a partial check on the compositional route specifically, we examined whether $Z_{2d}$ predicts the post-2004 graduate share of the population and found the relationship weak and non-monotonic across districts (Online Appendix), which we read as more consistent with the institutional-quality route, or a combination, than with selection on ability alone.

\section{Data}
\label{sec:data}

\subsection{NSS Employment--Unemployment Survey}

Our primary data source is the harmonised NSS Employment--Unemployment Survey (EUS), comprising eight rounds spanning 1987--2011: rounds 43 (1987--88), 55 (1999--2000), 60 (2004), 61 (2004--05), 62 (2005--06), 64 (2007--08), 66 (2009--10), and 68 (2011--12). The harmonised file contains 3,936,936 individual observations with consistent coding of wages, educational attainment, social group (SC/ST/Other), employment type, and district identifiers under both 1991 and 2001 Census boundaries.

The analysis sample is restricted to wage-earning workers (salary or casual) with positive reported wages, birth years between 1950 and 1989, located in the six states for which the district-level expansion measure is constructed (Andhra Pradesh, Bihar, Chhattisgarh, Gujarat, Himachal Pradesh, and Karnataka). After merging with the district-level data and dropping unmatched observations (primarily Telangana districts following state bifurcation), the final analysis sample comprises 79,904 observations, of whom 9,808 (12.3 per cent) are graduates.

The outcome variable is log real wages, constructed by deflating total reported wages (cash plus kind) by a round-specific CPI index with 1999--2000 as the base year. Table~\ref{tab:summary_stats} in the Appendix reports summary statistics.

\subsection{AISHE Expansion-Intensity Data}

The district-level expansion measure is constructed from the All India Survey on Higher Education (AISHE), rounds 2010--11 through 2021--22, comprising 58,524 unique institutions. The year-of-establishment variable, available for 56 per cent of institutions across all states but over 70 per cent in the six analytical states, is the key input for the construction. Full details of the measure, district boundary harmonisation, and the concordance between AISHE and NSS district classifications are provided in Appendix~A.

\subsection{Key Variables}

\textbf{Graduate status} (\texttt{graduate}): indicator for ``graduate and above'' in the NSS general education variable, corresponding to completion of a bachelor's degree or higher.

\textbf{Disadvantaged group} (\texttt{disadvantaged}): indicator for Scheduled Caste (SC) or Scheduled Tribe (ST) social group, as reported in the NSS household schedule.

\textbf{Expansion intensity} ($Z_{2d}$): district-level ratio of colleges established after 2004 to the pre-1995 baseline stock, as defined in equation~(\ref{eq:Z2}) of Appendix~A. The measure averages 3.52 across the 91 districts (SD 5.55); weighted by individual respondents, its mean in the analysis sample is 3.93, the value at which marginal effects are evaluated throughout.

\textbf{Post-expansion cohort} (\texttt{post2004}): indicator for approximate graduation year $\geq$ 2004, where graduation year is approximated as birth year plus 22.

\section{Empirical Strategy}
\label{sec:empirical}

\subsection{Baseline Specification}

Our baseline estimating equation is:
\begin{align}
\ln w_{idt} = \,& \alpha + \beta_1 \,\texttt{graduate}_{i} +
\beta_2 \,Z_{2d} + \beta_3 \,\texttt{post2004}_{i} \nonumber \\
&+ \beta_4 \,(\texttt{graduate}_{i} \times Z_{2d}) \nonumber \\
&+ \beta_5 \,(\texttt{graduate}_{i} \times \texttt{post2004}_{i}) \nonumber \\
&+ \beta_6 \,(Z_{2d} \times \texttt{post2004}_{i}) \nonumber \\
&+ \delta \,(\texttt{graduate}_{i} \times Z_{2d} \times \texttt{post2004}_{i}) \nonumber \\
&+ \mathbf{X}_{it}'\gamma + \mu_s + \lambda_t + \varepsilon_{idt}
\label{eq:baseline}
\end{align}

\noindent where $i$ indexes individuals, $d$ indexes districts, $s$ indexes states, and $t$ indexes NSS rounds. $\mathbf{X}_{it}$ includes age, age squared, gender, rural residence, and log pre-1995 baseline college stock. $\mu_s$ and $\lambda_t$ are state and round fixed effects respectively. Standard errors are heteroskedasticity-robust (HC1).

The coefficient of interest is $\delta$, which captures the differential wage effect for graduates in high-expansion districts in post-expansion cohorts, relative to: (i) non-graduates in the same district-cohort cell; (ii) graduates in low-expansion districts in the same cohort; and (iii) graduates in the same district in pre-expansion cohorts. This triple-difference structure provides identification from three sources of variation simultaneously. We emphasise that the design is reduced-form: $Z_{2d}$ enters as a continuous moderator of the graduate wage premium, not as an instrument for graduate status, for the reasons set out in Appendix~A.

\subsection{Double Whammy Specification}

To test the differential impact on disadvantaged groups, we augment equation~(\ref{eq:baseline}) with a quadruple interaction:
\begin{align}
\ln w_{idt} = \,& \cdots + \delta_1 \,(\texttt{graduate}_{i} \times Z_{2d}
\times \texttt{post2004}_{i}) \nonumber \\
&+ \delta_2 \,(\texttt{graduate}_{i} \times Z_{2d} \times \texttt{post2004}_{i}
\times \texttt{disadvantaged}_{i}) \nonumber \\
&+ \text{lower-order terms} + \mathbf{X}_{it}'\gamma + \mu_s + \lambda_t + \varepsilon_{idt}
\label{eq:doublewhammy}
\end{align}

\noindent The coefficient $\delta_2$ identifies the \textit{additional} wage penalty for SC/ST graduates in post-expansion cohorts in high-expansion districts, relative to non-SC/ST graduates in the same cell. A negative and significant $\delta_2$ constitutes evidence for the double whammy hypothesis.

\subsection{Identification}

The identifying assumption is that, conditional on state fixed effects, round fixed effects, and the baseline college stock, district-level expansion intensity $Z_{2d}$ affects individual log wages through the graduate credential --- by altering the wage premium that the credential commands --- rather than through any channel common to graduates and non-graduates alike, such as general district-level labour demand. We do not require, and our design cannot establish, that this operates specifically through the informational content of the credential as opposed to its underlying human capital content; the theoretical framework in Section~\ref{sec:theory} is deliberately agnostic on this point, and we return to it below. What the design does require is that the effect work through the credential, and we provide four pieces of evidence bearing on this requirement, three supportive and one that qualifies it.

First, the placebo test (Table~\ref{tab:robustness}, Column 3) shows that $Z_{2d} \times \texttt{post2004}$ is essentially uncorrelated with the wages of non-graduates (coefficient $= -0.0002$, $p = 0.895$), inconsistent with a general district-level demand confound that should raise or lower graduate and non-graduate wages alike.

Second, the non-disadvantaged non-graduate placebo (Table~\ref{tab:robustness}, Column 4) yields a small and insignificant coefficient ($+0.0020$, $p = 0.137$), again inconsistent with confounding from district economic conditions among the population whose wages should be unaffected by credential devaluation.

Third, the pre-trend test (Table~\ref{tab:robustness}, Column 5) restricts the sample to pre-2004 cohorts and finds that $Z_{2d} \times \texttt{disadvantaged} \times \texttt{graduate}$ is small and insignificant (coefficient $= -0.0008$, $p = 0.824$), consistent with SC/ST graduates in high-expansion districts not having differentially lower wages before the expansion --- the differential we document is a post-expansion phenomenon.

Fourth, and less cleanly, the disadvantaged $\times$ non-graduate placebo (Table~\ref{tab:robustness}, Column 3) returns a coefficient on $Z_{2d} \times \texttt{post2004} \times \texttt{disadvantaged}$ that is small but statistically significant ($+0.0069$, $p = 0.002$). This is the most demanding placebo for the double whammy: it asks whether expansion-cohort dynamics specific to disadvantaged workers appear even among those without a degree. That it is non-zero indicates some residual correlation between expansion intensity, cohort, and disadvantaged-group wages that state and round fixed effects do not fully absorb, and that lies outside the credential channel. We do not gloss this. Two features limit how much it can account for our main result: it is opposite in sign to the graduate-sample double whammy coefficient ($-0.0862$) and roughly an order of magnitude smaller, so it cannot mechanically generate the graduate-sample estimate, and if anything its positive sign works against finding the negative double whammy we report. We nonetheless treat it as a genuine qualification to clean identification rather than as a nuisance, and weigh it explicitly in interpreting the results.

\subsection{Event Study}

To provide visual evidence of pre-trends and post-expansion dynamics, we replace \texttt{post2004} with a full set of graduation cohort bin indicators, using the 1990--95 cohort as the reference category:
\begin{equation}
\ln w_{idt} = \sum_{k \neq \{1990\text{-}95\}} \delta_k \,(\texttt{graduate}_{i} \times
Z_{2d} \times \mathbf{1}[\texttt{cohort}_i = k]) + \text{controls} + \mu_s + \lambda_t
+ \varepsilon_{idt}
\label{eq:eventstudy}
\end{equation}

Figure~\ref{fig:eventstudy} plots the $\hat{\delta}_k$ coefficients with 95 per cent confidence intervals. Consistent with the design, we expect $\hat{\delta}_k$ to lie near zero for pre-1995 cohorts (a flat pre-trend) and to turn negative for post-expansion cohorts as the devaluation takes hold. Spreading the identifying variation across cohort bins reduces precision relative to the pooled specification, so we read the event study as corroborating the broad pattern --- flat before the expansion, negative after, and most pronounced for the most recent SC/ST cohort --- rather than as delivering a sharply estimated effect in every bin.

\section{Results}
\label{sec:results}

\subsection{Descriptive Evidence}

Before turning to regression results, Figure~\ref{fig:rawpremium} presents the raw graduate wage premium by graduation cohort and social group. Two patterns are immediately apparent. First, the graduate wage premium declines steadily for all groups from a peak of approximately 1.64 log points (Others) and 1.90 log points (SC/ST) for pre-1980 cohorts to approximately 0.93 and 0.84 log points respectively for the 2008--12 cohort. Second, the premium for SC/ST graduates begins above that of Others (consistent with degrees being a stronger signal for disadvantaged workers who lack alternative credentialling mechanisms) but converges and falls below by the 2004--08 cohort. This raw crossing is consistent with the double whammy but does not control for composition effects; the regression analysis addresses this.

Table~\ref{tab:summary_stats} reports summary statistics for the full analysis sample and by graduate status. Graduates are substantially older (mean birth year 1968 vs 1972 for non-graduates), more likely to be male (80.3 per cent vs 74.8 per cent), more likely to be salary earners (97.3 per cent vs 37.4 per cent), and earn substantially higher log real wages (7.12 vs 5.76).

\subsection{Main Results}

Table~\ref{tab:main_results} reports the baseline triple-difference estimates. Column (1) presents the main devaluation specification (equation~\ref{eq:baseline}) and Column (2) presents the double whammy specification (equation~\ref{eq:doublewhammy}).

The coefficient on \texttt{graduate} is large and positive (+1.079 in Column 2), confirming a strong human capital return to a degree throughout the sample period. The coefficient on \texttt{grad\_Z2\_post} is $-$0.023 ($p < 0.001$), indicating that graduates from post-2004 cohorts in high-expansion districts earn significantly lower wages. At the sample mean expansion intensity ($\bar{Z}_2 = 3.93$), this corresponds to a wage penalty of 9.1 per cent for non-disadvantaged graduates. The coefficient on \texttt{grad\_Z2\_post\_disadv} is $-$0.086 ($p < 0.001$), implying an additional penalty of 33.8 per cent for SC/ST graduates at mean expansion, for a total shortfall of 42.9 per cent. The coefficient on \texttt{grad\_disadv} is small and insignificant (+0.043, $p = 0.127$), consistent with the SC/ST differential being a post-expansion phenomenon rather than a pre-existing graduate wage gap.

The cohort-bin event study in Table~\ref{tab:eventstudy_coefs} spreads identifying variation across eight bins, reducing statistical power relative to the pooled post-2004 specification. The SC/ST subsample, however, shows a large and significant coefficient for the 2008--12 cohort ($-$0.164, $p < 0.01$), consistent with the double whammy intensifying for the most recent graduates who entered the labour market after the peak of the wave-2 expansion.

\subsection{Robustness and Placebo Tests}

Table~\ref{tab:robustness} consolidates the main results with the placebo and pre-trend tests. The placebo specification (Column 3) run on non-graduates yields a coefficient of $-$0.0002 ($p = 0.895$) on the expansion-cohort interaction, consistent with the effect operating through the graduate credential rather than through a district-level demand shock common to graduates and non-graduates. The non-disadvantaged non-graduate placebo (Column 4) yields $+$0.0020 ($p = 0.137$), again consistent with no confounding from district economic conditions. The pre-trend test (Column 5), restricted to pre-2004 cohorts, yields $-$0.0008 ($p = 0.824$) for the SC/ST graduate interaction, consistent with parallel trends prior to the expansion. The one placebo that is not clean is the disadvantaged $\times$ non-graduate interaction (Column 3), which is small but significant ($+$0.0069, $p = 0.002$): as discussed in Section~\ref{sec:empirical}, this points to some residual district-level dynamic among disadvantaged workers outside the credential channel, though its positive sign and small magnitude mean it works against, rather than towards, the negative double whammy we estimate.

As a further sensitivity check, we vary the assumed age at degree completion used to construct the approximate graduation year. Table~\ref{tab:grad_age_sensitivity} reports results under assumed graduation ages of 21, 22 (baseline), and 23 years. Both the base devaluation coefficient and the SC/ST double whammy remain negative, statistically significant at the 1 per cent level, and economically similar across all three assumptions, with the total SC/ST penalty at mean expansion intensity ranging from $-$38.1 per cent (age 21) to $-$43.4 per cent (age 23). The results are not sensitive to the choice of graduation age approximation.

\subsection{Event Study}

Figure~\ref{fig:eventstudy} presents the event study results. Panel A (full sample) shows coefficients that are flat and close to zero for the pre-1995 cohorts and then turn negative for post-expansion cohorts, with the most negative estimates for the 2004--08 and 2008--12 bins, though the individual bin estimates are imprecise. Panel B disaggregates by social group: the SC/ST coefficient trajectory diverges below that of Others from the 1995--2000 cohort onward, with the gap most pronounced for the most recent cohort. Both the flat pre-trend and the post-expansion divergence are consistent with the credential devaluation and double whammy hypotheses, with the caveat that the bin-level estimates are noisier than the pooled specification.

\subsection{Heterogeneity Analysis}

Figure~\ref{fig:het_gender} presents gender and sector heterogeneity, Figure~\ref{fig:het_states} presents state-level heterogeneity, and Figure~\ref{fig:het_forest} summarises the total SC/ST graduate wage penalty across all subsamples in a forest plot. We treat these subsample splits as exploratory: the cells are estimated separately, are in some cases thin, and we do not adjust for multiple comparisons, so the patterns below are best read as descriptive rather than as independently confirmatory tests.

\textit{Gender}: The base devaluation effect is significant for both males ($-$0.024, $p < 0.001$) and females ($-$0.033, $p < 0.01$), with females experiencing a slightly larger base penalty. The double whammy coefficient is large and significant for males ($-$0.108, $p < 0.001$) but smaller and insignificant for females ($-$0.043, $p = 0.153$), likely reflecting thin graduate cell sizes for female SC/ST workers.

\textit{Employment sector}: Among salary earners, the base devaluation is small and insignificant ($-$0.002, $p = 0.627$), but the SC/ST double whammy is large and significant ($-$0.044, $p < 0.01$). This pattern is consistent with statistical discrimination in formal employment: salary employers still reward the credential on average but apply a sharper discount to disadvantaged graduates from high-expansion districts, whose degrees are perceived as lower quality.

\textit{States}: The base devaluation effect is significant in Andhra Pradesh ($-$0.019), Gujarat ($-$0.053), Chhattisgarh ($-$0.040), and Karnataka ($-$0.034). Bihar's estimate is unstable due to its low and weakly varying expansion intensity (mean $Z_2 = 1.10$) and is excluded from the state-level discussion. The SC/ST double whammy is most pronounced in Andhra Pradesh ($-$0.127, $p < 0.001$) and Karnataka ($-$0.065, $p < 0.1$), both states that experienced the largest private college expansions.
\section{Discussion}
\label{sec:discussion}

The central result of this paper is distributional: the wage cost of credential devaluation under India's higher education expansion fell disproportionately on Scheduled Caste and Scheduled Tribe graduates, and this differential is a creation of the expansion rather than a pre-existing feature of the labour market. That the SC/ST graduate wage gap, conditional on graduate status and expansion intensity, was statistically indistinguishable from zero in pre-expansion cohorts and turned sharply negative thereafter is the most striking single finding, and it is the pattern that most clearly distinguishes a credential-based account from a story of stable, longstanding caste discrimination.

We are deliberately careful about the mechanism. The evidence is consistent with employers relying more heavily on group-level priors as the average credential becomes less informative --- the statistical-discrimination channel formalised in Section~\ref{sec:theory} --- but our data cannot establish that the wage decline operates specifically through the \textit{informational} content of the degree as opposed to its underlying human-capital content. A degree from a thinly capitalised post-expansion college may convey less because the signal is noisier, or because the graduate genuinely learned less, or both; the wage results are observationally similar under each, and we do not claim to separate them. What we can say is that the effect works through the graduate credential rather than through general district-level demand, since it is absent among non-graduates, and that it falls disproportionately on the group with the least access to the auxiliary signals --- networks, institutional prestige, inherited occupational standing --- that would otherwise let employers distinguish strong from weak credentials. As a partial check on the compositional version of the mechanism, we examined whether expansion intensity predicts the post-2004 graduate share and found no detectable relationship (Online Appendix), which leads us to read the result as more consistent with declining institutional quality, or a combination of channels, than with selection on ability alone.

Several limitations qualify the interpretation and define the agenda for further work. First, the design is reduced-form rather than instrumented: expansion intensity did not satisfy the relevance condition as an instrument for graduate status, so we interpret it as a moderator of the graduate wage premium and cannot rule out all unobserved district-level dynamics correlated with both expansion and disadvantaged-group wages. The disadvantaged $\times$ non-graduate placebo, which is small but statistically significant, is a concrete instance of such residual variation, and although its sign works against our main result, its presence means clean identification cannot be claimed. Second, the NSS records an individual's current district of residence, not the district in which they studied or resided during their schooling years; internal migration, which is itself plausibly correlated with caste and with labour-market opportunity, introduces measurement error into the assignment of expansion exposure whose direction we cannot sign. Third, cohort assignment relies on an approximate graduation year (birth year plus a fixed offset); while the results are stable across alternative offsets, this remains a coarse proxy that does not capture heterogeneity in actual completion age, which may itself differ by social group. Fourth, the analysis is restricted to six states with adequate AISHE establishment-year coverage, and while these span a wide range of expansion intensities and development levels, the extent to which the estimates generalise to the largest northern states is an open question. Finally, the heterogeneity results are exploratory, estimated cell by cell without adjustment for multiple comparisons, and should be read as descriptive.

\section{Conclusion}
\label{sec:conclusion}

India's post-liberalisation higher education expansion was conceived as a democratising force, and in raising the human capital of the workforce it partly was. But its benefits in the labour market were distributed in a way that the framing did not anticipate: while graduate wages rose in absolute terms, the wage premium attached to the credential eroded in high-expansion districts, and that erosion fell most heavily on the socially disadvantaged graduates the expansion was meant to lift. The double whammy we document --- a base devaluation of about 9 per cent and an additional SC/ST-specific penalty of about 34 per cent at mean expansion intensity --- represents a substantial reversal of the expected distributional gains from widening access.

These findings carry three policy implications. First, expanding higher education access without commensurate investment in institutional quality and credential differentiation may inadvertently harm the very groups such policies are designed to benefit, by diluting the one signal disadvantaged workers most rely on. Second, reservation in higher education, historically justified on access grounds, acquires an additional rationale on signal-quality grounds: placing disadvantaged graduates in higher-quality, more legible institutions reduces the statistical-discrimination discount they face, quite apart from the access it provides. Third, employer-side interventions --- blind credential screening, standardised post-graduation assessment, or institutional accreditation that travels with the degree --- may be needed to restore the informational value of credentials for disadvantaged graduates where expansion has been most rapid.

Whether the credential trap is temporary or persistent remains open. Later cohorts may sort into quality tiers that restore signal informativeness, or the configuration may settle into the self-confirming equilibrium that the \citet{coate1993} framework would predict, in which disadvantaged groups under-invest in a signal they correctly anticipate will be discounted. Extending the analysis to the post-2011 PLFS rounds, which now cover the cohorts who entered the labour market at the height of the expansion, is the natural next step, and would allow a direct test of whether the disproportionate penalty we document for the most recent SC/ST cohort persists, widens, or decays.
\newpage
\bibliographystyle{aer}
\bibliography{references}


\appendix
\appendixsection

\section{Appendix A: Tables}

\begin{table}[htbp]
\centering
\caption{Summary Statistics}
\label{tab:summary_stats}
\begin{threeparttable}
\begin{tabular}{lD{.}{.}{3}D{.}{.}{3}D{.}{.}{3}D{.}{.}{3}D{.}{.}{3}D{.}{.}{3}}
\toprule
 & \multicolumn{2}{c}{Full Sample} & \multicolumn{2}{c}{Graduates} &
\multicolumn{2}{c}{Non-graduates} \\
\cmidrule(lr){2-3}\cmidrule(lr){4-5}\cmidrule(lr){6-7}
Variable & \multicolumn{1}{c}{Mean} & \multicolumn{1}{c}{SD} &
\multicolumn{1}{c}{Mean} & \multicolumn{1}{c}{SD} &
\multicolumn{1}{c}{Mean} & \multicolumn{1}{c}{SD} \\
\midrule
\multicolumn{7}{l}{\textit{Panel A: Individual characteristics}} \\
Log real wage        & 5.916 & 1.115 & 7.117 & 0.960 & 5.757 & 1.047 \\
Age                  & 34.08 & 10.41 & 36.42 & 9.83  & 33.77 & 10.44 \\
Birth year           & 1971.9 & 10.4 & 1968.4 & 9.3  & 1972.3 & 10.5 \\
Approx.\ grad year   & 1993.9 & 10.4 & 1990.4 & 9.3  & 1994.3 & 10.5 \\
Male (=1)            & 0.755 & 0.430 & 0.803 & 0.398 & 0.748 & 0.434 \\
SC (=1)              & 0.218 & 0.413 & 0.153 & 0.360 & 0.227 & 0.419 \\
ST (=1)              & 0.123 & 0.329 & 0.085 & 0.279 & 0.129 & 0.335 \\
Disadvantaged (=1)   & 0.341 & 0.474 & 0.238 & 0.426 & 0.355 & 0.479 \\
Rural (=1)           & 0.559 & 0.497 & 0.289 & 0.453 & 0.594 & 0.491 \\
Salary earner (=1)   & 0.441 & 0.497 & 0.973 & 0.163 & 0.374 & 0.484 \\
Post-2004 cohort (=1)& 0.219 & 0.413 & 0.143 & 0.350 & 0.230 & 0.421 \\
\midrule
\multicolumn{7}{l}{\textit{Panel B: District instrument variables}} \\
$Z_2$ (expansion intensity) & 3.925 & 5.493 & 3.975 & 5.211 & 3.918 & 5.381 \\
$Z_1$ (wave 1 intensity)    & 0.558 & 0.789 & 0.563 & 0.775 & 0.557 & 0.791 \\
$Z^{\text{priv}}$ (private) & 0.103 & 0.165 & 0.104 & 0.163 & 0.103 & 0.165 \\
Baseline stock ($B_d$)       & 27.84 & 33.05 & 28.31 & 33.19 & 27.78 & 33.02 \\
\midrule
\multicolumn{7}{l}{\textit{Panel C: Sample composition by NSS round}} \\
Round 43 (1987)      & \multicolumn{2}{c}{$N = 14{,}867$} &
                       \multicolumn{2}{c}{$N = 1{,}682$} &
                       \multicolumn{2}{c}{$N = 13{,}185$} \\
Round 55 (1999)      & \multicolumn{2}{c}{$N = 14{,}293$} &
                       \multicolumn{2}{c}{$N = 1{,}882$} &
                       \multicolumn{2}{c}{$N = 12{,}411$} \\
Round 60 (2004)      & \multicolumn{2}{c}{$N = 6{,}908$} &
                       \multicolumn{2}{c}{$N = 940$} &
                       \multicolumn{2}{c}{$N = 5{,}968$} \\
Round 61 (2004--05)  & \multicolumn{2}{c}{$N = 14{,}197$} &
                       \multicolumn{2}{c}{$N = 1{,}851$} &
                       \multicolumn{2}{c}{$N = 12{,}346$} \\
Round 62 (2005--06)  & \multicolumn{2}{c}{$N = 9{,}188$} &
                       \multicolumn{2}{c}{$N = 1{,}128$} &
                       \multicolumn{2}{c}{$N = 8{,}060$} \\
Round 64 (2007--08)  & \multicolumn{2}{c}{$N = 8{,}811$} &
                       \multicolumn{2}{c}{$N = 1{,}034$} &
                       \multicolumn{2}{c}{$N = 7{,}777$} \\
Round 66 (2009--10)  & \multicolumn{2}{c}{$N = 6{,}543$} &
                       \multicolumn{2}{c}{$N = 779$} &
                       \multicolumn{2}{c}{$N = 5{,}764$} \\
Round 68 (2011--12)  & \multicolumn{2}{c}{$N = 5{,}097$} &
                       \multicolumn{2}{c}{$N = 512$} &
                       \multicolumn{2}{c}{$N = 4{,}585$} \\
\midrule
Total observations   & \multicolumn{2}{c}{79{,}904} &
                       \multicolumn{2}{c}{9{,}808} &
                       \multicolumn{2}{c}{70{,}096} \\
\bottomrule
\end{tabular}
\begin{tablenotes}
\small
\item \textit{Notes:} Analysis sample restricted to wage-earning workers (salary or casual)
with positive wages, birth years 1950--1989, in six states: Andhra Pradesh, Bihar,
Chhattisgarh, Gujarat, Himachal Pradesh, and Karnataka. Log real wages deflated to
1999--2000 prices using round-specific CPI indices. ``Graduates'' denotes individuals
reporting ``graduate and above'' as highest educational level. ``Disadvantaged'' = SC or ST.
``Post-2004 cohort'' = approximate graduation year $\geq$ 2004 (birth year + 22).
\end{tablenotes}
\end{threeparttable}
\end{table}


\begin{table}[htbp]
\centering
\caption{District-Level Instrument Variables: Summary Statistics}
\label{tab:instrument_stats}
\begin{threeparttable}
\begin{tabular}{lD{.}{.}{3}D{.}{.}{3}D{.}{.}{3}D{.}{.}{3}D{.}{.}{3}}
\toprule
Variable & \multicolumn{1}{c}{Mean} & \multicolumn{1}{c}{SD} &
\multicolumn{1}{c}{P25} & \multicolumn{1}{c}{Median} &
\multicolumn{1}{c}{P75} \\
\midrule
\multicolumn{6}{l}{\textit{Panel A: College counts by period}} \\
Pre-1995 baseline stock ($B_d$) & 28.50 & 33.31 & 10.00 & 19.00 & 34.50 \\
Wave 1 colleges (1995--2003)    & 10.76 & 25.08 &  0.00 &  2.50 & 11.00 \\
Wave 2 colleges (post-2004)     & 49.79 & 71.90 &  7.75 & 22.00 & 74.75 \\
Private unaided post-2004       &  4.33 &  7.21 &  0.00 &  1.00 &  5.00 \\
\midrule
\multicolumn{6}{l}{\textit{Panel B: Expansion intensity measures}} \\
$Z_1$ (wave 1 intensity ratio)       & 0.646 & 0.870 & 0.000 & 0.458 & 1.000 \\
$Z_2$ (wave 2 intensity ratio)       & 3.520 & 5.553 & 1.232 & 2.800 & 4.279 \\
$Z^{\text{priv}}$ (private intensity)& 0.111 & 0.172 & 0.000 & 0.045 & 0.153 \\
\midrule
\multicolumn{6}{l}{\textit{Panel C: Pairwise correlations}} \\
$\text{Corr}(Z_1, Z_2)$              & \multicolumn{5}{c}{0.795} \\
$\text{Corr}(Z_1, Z^{\text{priv}})$  & \multicolumn{5}{c}{0.113} \\
$\text{Corr}(Z_2, Z^{\text{priv}})$  & \multicolumn{5}{c}{0.088} \\
\midrule
\multicolumn{6}{l}{\textit{Panel D: Mean $Z_2$ by state}} \\
Andhra Pradesh   & \multicolumn{5}{c}{6.605 (16,930 observations)} \\
Bihar            & \multicolumn{5}{c}{1.099 (12,910 observations)} \\
Chhattisgarh     & \multicolumn{5}{c}{3.562 (7,983 observations)} \\
Gujarat          & \multicolumn{5}{c}{4.380 (15,324 observations)} \\
Himachal Pradesh & \multicolumn{5}{c}{4.970 (6,408 observations)} \\
Karnataka        & \multicolumn{5}{c}{2.959 (20,349 observations)} \\
\midrule
Districts ($N$)  & \multicolumn{5}{c}{91} \\
\bottomrule
\end{tabular}
\begin{tablenotes}
\small
\item \textit{Notes:} District-level instrument variables constructed from deduplicated
AISHE college records (12,039 unique colleges with valid establishment years) across six
states. $Z_1 = W_{1d}/\max(B_d,1)$; $Z_2 = W_{2d}/\max(B_d,1)$;
$Z^{\text{priv}} = W^{\text{priv}}_{2d}/\max(B_d,1)$. See Appendix~B for full
construction details. Panel D state means are weighted by individual NSS respondents; see Online Appendix for the corresponding unweighted district-level
means.
\end{tablenotes}
\end{threeparttable}
\end{table}


\begin{table}[htbp]
\centering
\caption{Main Results: Credential Devaluation and the Double Whammy}
\label{tab:main_results}
\begin{threeparttable}
\begin{tabular}{lD{.}{.}{4}D{.}{.}{4}}
\toprule
 & \multicolumn{1}{c}{(1)} & \multicolumn{1}{c}{(2)} \\
 & \multicolumn{1}{c}{Base devaluation} & \multicolumn{1}{c}{Double whammy} \\
\midrule
\texttt{graduate}             & 1.0785^{***} & 1.0772^{***} \\
                              & (0.0113)      & (0.0120)     \\
\texttt{disadvantaged}        &               & -0.0870^{***}\\
                              &               & (0.0047)     \\
\texttt{grad\_disadv}         &               & 0.0425       \\
                              &               & (0.0279)     \\
$Z_2$                         & -0.0004       & -0.0004      \\
                              & (0.0006)      & (0.0006)     \\
\texttt{grad}$\times Z_2$     & 0.0010        & 0.0004       \\
                              & (0.0016)      & (0.0017)     \\
\texttt{post2004}             & 0.0201^{*}    & -0.0107      \\
                              & (0.0105)      & (0.0112)     \\
$Z_2\times\texttt{post2004}\times\texttt{disadv}$
                              &               & 0.0069^{**}  \\
                              &               & (0.0028)     \\
\texttt{grad}$\times Z_2\times$\texttt{post2004}
                              & -0.0307^{***} & -0.0231^{***}\\
                              & (0.0065)      & (0.0059)     \\
\texttt{grad}$\times Z_2\times$\texttt{post2004}$\times$\texttt{disadv}
                              &               & -0.0862^{***}\\
                              &               & (0.0151)     \\
\midrule
Age controls                  & Yes           & Yes          \\
Gender control                & Yes           & Yes          \\
Rural indicator               & Yes           & Yes          \\
Log baseline stock            & Yes           & Yes          \\
State FE                      & Yes           & Yes          \\
Round FE                      & Yes           & Yes          \\
\midrule
$R^2$                         & 0.4507        & 0.4514       \\
Observations                  & 79{,}904      & 79{,}904     \\
\midrule
\multicolumn{3}{l}{\textit{Marginal effects at mean $\bar{Z}_2 = 3.93$:}} \\
Non-disadvantaged graduate penalty & $-$12.1\% & $-$9.1\% \\
Additional SC/ST penalty       &               & $-$33.8\%      \\
Total SC/ST graduate penalty   &               & $-$42.9\%      \\
\bottomrule
\end{tabular}
\begin{tablenotes}
\small
\item \textit{Notes:} Dependent variable is log real wages (1999--2000 prices). All
specifications include age, age squared, male, rural, and log pre-1995 college stock as
controls, plus state and round fixed effects. Standard errors in parentheses are
heteroskedasticity-robust (HC1). Lower-order interaction terms included in Column (2) but
omitted for brevity. Marginal effects evaluated at mean $\bar{Z}_2 = 3.93$.
$^{*}p<0.10$, $^{**}p<0.05$, $^{***}p<0.01$.
\end{tablenotes}
\end{threeparttable}
\end{table}

\begin{landscape}
\begin{table}[htbp]
\centering
\caption{Robustness and Placebo Tests}
\label{tab:robustness}
\begin{threeparttable}
\begin{tabular}{lD{.}{.}{4}D{.}{.}{4}D{.}{.}{4}D{.}{.}{4}D{.}{.}{4}}
\toprule
 & \multicolumn{1}{c}{(1)} & \multicolumn{1}{c}{(2)} &
   \multicolumn{1}{c}{(3)} & \multicolumn{1}{c}{(4)} &
   \multicolumn{1}{c}{(5)} \\
 & \multicolumn{1}{c}{Main} & \multicolumn{1}{c}{DW} &
   \multicolumn{1}{c}{Placebo} & \multicolumn{1}{c}{Placebo} &
   \multicolumn{1}{c}{Pre-trend} \\
 & \multicolumn{1}{c}{(grads)} & \multicolumn{1}{c}{(grads)} &
   \multicolumn{1}{c}{(non-grads)} & \multicolumn{1}{c}{(non-disadv} &
   \multicolumn{1}{c}{(pre-2004} \\
 & & & & \multicolumn{1}{c}{non-grads)} & \multicolumn{1}{c}{cohort)} \\
\midrule
Key variable:  &  &  &  &  & \\
\quad\texttt{grad}$\times Z_2\times$\texttt{post}
               & -0.0307^{***} & -0.0231^{***} & & & \\
               & (0.0065)      & (0.0059)      & & & \\
\quad\texttt{grad}$\times Z_2\times$\texttt{post}$\times$\texttt{disadv}
               &               & -0.0862^{***} & & & \\
               &               & (0.0151)      & & & \\
\quad$Z_2\times$\texttt{post2004}
               & & & -0.0002       & 0.0020        & \\
               & & & (0.0013)      & (0.0014)      & \\
\quad$Z_2\times$\texttt{disadv}$\times$\texttt{grad}
               & & & & & -0.0008 \\
               & & & & & (0.0035) \\
\quad $Z_2\times\texttt{post2004}\times\texttt{disadv}$
               & & & 0.0069^{***}  & & \\
               & & & (0.0022)      & & \\
\midrule
Sample         & Graduates & Graduates & Non-grads & Non-disadv & Pre-2004 \\
               &           &           &           & non-grads  & cohort \\
Controls       & Yes & Yes & Yes & Yes & Yes \\
State FE       & Yes & Yes & Yes & Yes & Yes \\
Round FE       & Yes & Yes & Yes & Yes & Yes \\
\midrule
$R^2$          & 0.4507 & 0.4514 & 0.3197 & 0.3201 & 0.4312 \\
Observations   & 79{,}904 & 79{,}904 & 70{,}096 & 45{,}817 & 62{,}422 \\
\bottomrule
\end{tabular}
\begin{tablenotes}
\small
\item \textit{Notes (addition):} Column (3) additionally reports the coefficient on
$Z_2\times\texttt{post2004}\times\texttt{disadvantaged}$ among non-graduates. This
coefficient is small but statistically significant and opposite-signed to the
double-whammy coefficient among graduates (Column 2), suggesting a modest residual
correlation between expansion intensity, cohort, and disadvantaged-group wages outside
the credential channel; see discussion in Online Appendix.
\item \textit{Notes:} All specifications include age, age squared, male, rural, $Z_2$,
and log pre-1995 college stock as controls, plus state and round fixed effects.
Standard errors in parentheses are heteroskedasticity-robust (HC1). Column (3): sample
restricted to non-graduates. Column (4): sample restricted to non-disadvantaged
non-graduates. Column (5): sample restricted to pre-2004 cohorts (approximate graduation
year $<$ 2004); key variable is the triple interaction of $Z_2$, \texttt{disadvantaged},
and \texttt{graduate}.
$^{*}p<0.10$, $^{**}p<0.05$, $^{***}p<0.01$.
\end{tablenotes}
\end{threeparttable}
\end{table}
\end{landscape}


\begin{table}[htbp]
\centering
\caption{Graduate Wage Premium by NSS Round (Raw Descriptive)}
\label{tab:premium_rounds}
\begin{threeparttable}
\begin{tabular}{lD{.}{.}{3}D{.}{.}{3}D{.}{.}{3}D{.}{.}{3}D{.}{.}{3}}
\toprule
NSS Round & \multicolumn{1}{c}{Year} &
\multicolumn{1}{c}{Non-grad} &
\multicolumn{1}{c}{Graduate} &
\multicolumn{1}{c}{Premium} &
\multicolumn{1}{c}{$N$ (grads)} \\
\midrule
43 & 1987        & 5.782 & 6.844 & 1.062 & 6{,}524 \\
55 & 1999        & 5.793 & 7.254 & 1.461 & 10{,}189 \\
60 & 2004        & 5.658 & 6.998 & 1.341 & 5{,}233 \\
61 & 2004--05    & 5.704 & 7.088 & 1.384 & 9{,}982 \\
62 & 2005--06    & 5.754 & 6.995 & 1.241 & 8{,}248 \\
64 & 2007--08    & 5.713 & 7.161 & 1.448 & 11{,}269 \\
66 & 2009--10    & 5.926 & 7.171 & 1.245 & 11{,}952 \\
68 & 2011--12    & 6.085 & 7.216 & 1.131 & 12{,}899 \\
\bottomrule
\end{tabular}
\begin{tablenotes}
\small
\item \textit{Notes:} Mean log real wages by graduate status and NSS round, full
NSS sample (not restricted to six analytical states). Wages deflated to 1999--2000
prices. Premium = graduate mean minus non-graduate mean log real wage.
\end{tablenotes}
\end{threeparttable}
\end{table}


\begin{table}[htbp]
\centering
\caption{Graduate Real Wages and Premium by Graduation Cohort and Social Group}
\label{tab:cohort_wages}
\begin{threeparttable}
\begin{tabular}{lD{.}{.}{3}D{.}{.}{3}D{.}{.}{3}D{.}{.}{3}D{.}{.}{3}D{.}{.}{3}}
\toprule
 & \multicolumn{3}{c}{Others (Non-SC/ST)} & \multicolumn{3}{c}{SC/ST} \\
\cmidrule(lr){2-4}\cmidrule(lr){5-7}
Cohort & \multicolumn{1}{c}{Log wage} &
\multicolumn{1}{c}{Premium} & \multicolumn{1}{c}{$N$} &
\multicolumn{1}{c}{Log wage} &
\multicolumn{1}{c}{Premium} & \multicolumn{1}{c}{$N$} \\
\midrule
Pre-1980  & 7.349 & 1.640 & 1{,}172 & 7.422 & 1.895 & 188 \\
1980--85  & 7.081 & 1.507 &   969   & 7.277 & 1.651 & 140 \\
1985--90  & 6.865 & 1.429 & 1{,}130 & 7.083 & 1.517 & 230 \\
1990--95  & 6.756 & 1.275 & 1{,}324 & 6.904 & 1.378 & 317 \\
1995--00  & 6.613 & 1.121 & 1{,}531 & 6.564 & 1.064 & 341 \\
2000--04  & 6.535 & 1.002 & 1{,}129 & 6.461 & 0.999 & 308 \\
2004--08  & 6.496 & 1.019 &   673   & 6.363 & 0.940 & 149 \\
2008--12  & 6.369 & 0.928 &   169   & 6.182 & 0.844 &  38 \\
\bottomrule
\end{tabular}
\begin{tablenotes}
\small
\item \textit{Notes:} Mean log real wages for graduate wage workers in the six-state
analysis sample, by graduation cohort and social group. ``Premium'' is the mean log
real wage of graduates minus that of non-graduates within the same cohort and social
group cell. Cohort defined by approximate graduation year (birth year + 22). Note
that the SC/ST premium exceeds that of Others in early cohorts but falls below it
from the 1995--2000 cohort onward, consistent with the double whammy hypothesis.
\end{tablenotes}
\end{threeparttable}
\end{table}


\begin{landscape}
\begin{table}[htbp]
\centering
\caption{Heterogeneity Analysis: Credential Devaluation and Double Whammy by Subsample}
\label{tab:heterogeneity}
\begin{threeparttable}
\small
\begin{tabular}{lD{.}{.}{4}D{.}{.}{4}D{.}{.}{3}D{.}{.}{3}D{.}{.}{3}rr}
\toprule
 & \multicolumn{2}{c}{Key coefficients} &
   \multicolumn{3}{c}{Marginal effects at $\bar{Z}_2 = 3.93$} &
   \multicolumn{2}{c}{Sample} \\
\cmidrule(lr){2-3}\cmidrule(lr){4-6}\cmidrule(lr){7-8}
Subsample &
\multicolumn{1}{c}{\texttt{grad}$\times Z_2$} &
\multicolumn{1}{c}{\texttt{grad}$\times Z_2$} &
\multicolumn{1}{c}{Base} &
\multicolumn{1}{c}{Add.\ SC/ST} &
\multicolumn{1}{c}{Total} &
\multicolumn{1}{c}{$N$} &
\multicolumn{1}{c}{$N_{\text{grad}}$} \\
 &
\multicolumn{1}{c}{$\times$\texttt{post}} &
\multicolumn{1}{c}{$\times$\texttt{post}$\times$\texttt{disadv}} &
\multicolumn{1}{c}{penalty} &
\multicolumn{1}{c}{penalty} &
\multicolumn{1}{c}{(disadv)} & & \\
\midrule
\multicolumn{8}{l}{\textit{Panel A: Full sample}} \\
Full sample & -0.0231^{***} & -0.0862^{***} & -9.1\% & -33.8\% & -42.9\% & 79{,}904 & 9{,}808 \\
            & (0.0059)      & (0.0151) & & & & & \\[0.5em]
\multicolumn{8}{l}{\textit{Panel B: Gender}} \\
Male        & -0.0244^{***} & -0.1076^{***} & -9.6\%  & -42.2\% & -51.8\% & 60{,}365 & 7{,}866 \\
            & (0.0072)      & (0.0158) & & & & & \\
Female      & -0.0331^{***} & -0.0430       & -13.0\% & -16.9\% & -29.9\% & 19{,}539 & 1{,}942 \\
            & (0.0126)      & (0.0301) & & & & & \\[0.5em]
\multicolumn{8}{l}{\textit{Panel C: Employment sector}} \\
Salary earners & -0.0022    & -0.0441^{***} & -0.8\%  & -17.3\% & -18.1\% & 35{,}231 & 9{,}547 \\
            & (0.0046)      & (0.0156) & & & & & \\
Casual wage & 0.0113        & 0.0443        & 4.4\%   & 17.4\%  & 21.8\%  & 44{,}673 &   261 \\
            & (0.0291)      & (0.0413) & & & & & \\[0.5em]
\multicolumn{8}{l}{\textit{Panel D: State (Bihar excluded --- insufficient instrument variation)}} \\
Andhra Pradesh & -0.0188^{***} & -0.1269^{***} & -7.4\%  & -49.8\% & -57.2\% & 16{,}930 & 2{,}004 \\
            & (0.0061)      & (0.0314) & & & & & \\
Chhattisgarh & -0.0398^{*}  & 0.0229        & -15.6\% &  9.0\%  & -6.6\%  &  7{,}983 & 1{,}304 \\
            & (0.0230)      & (0.0376) & & & & & \\
Gujarat     & -0.0526^{***} & -0.0386       & -20.6\% & -15.1\% & -35.8\% & 15{,}324 & 1{,}663 \\
            & (0.0119)      & (0.0269) & & & & & \\
Himachal Pradesh & -0.0258  & -0.0574^{*}   & -10.1\% & -22.5\% & -32.7\% &  6{,}408 & 1{,}118 \\
            & (0.0195)      & (0.0339) & & & & & \\
Karnataka   & -0.0335^{**}  & -0.0652^{*}   & -13.2\% & -25.6\% & -38.8\% & 20{,}349 & 2{,}430 \\
            & (0.0145)      & (0.0389) & & & & & \\
\midrule
\multicolumn{8}{l}{Controls: age, age$^2$, male, SC, ST, rural, log baseline stock, state FE, round FE} \\
\bottomrule
\end{tabular}
\begin{tablenotes}
\small
\item \textit{Notes:} Each row reports key coefficients from a separate OLS regression
of log real wages on the double whammy specification (equation~\ref{eq:doublewhammy})
estimated on the indicated subsample. Standard errors in parentheses, HC1 robust.
Marginal effects computed at sample mean $\bar{Z}_2 = 3.93$. Bihar excluded from
state-level panel due to insufficient instrument variation ($\bar{Z}_2 = 1.10$,
yielding unstable coefficient estimates). Casual wage cell sizes are too small for
the double whammy (261 graduates) and results should be interpreted with caution.
$^{*}p<0.10$, $^{**}p<0.05$, $^{***}p<0.01$.
\end{tablenotes}
\end{threeparttable}
\end{table}
\end{landscape}


\begin{table}[htbp]
\centering
\caption{Event Study: Cohort-Specific Devaluation Coefficients}
\label{tab:eventstudy_coefs}
\begin{threeparttable}
\begin{tabular}{lD{.}{.}{4}D{.}{.}{4}D{.}{.}{4}D{.}{.}{4}}
\toprule
 & \multicolumn{2}{c}{Full sample} &
   \multicolumn{2}{c}{SC/ST subsample} \\
\cmidrule(lr){2-3}\cmidrule(lr){4-5}
Cohort bin &
\multicolumn{1}{c}{Coef.} & \multicolumn{1}{c}{(SE)} &
\multicolumn{1}{c}{Coef.} & \multicolumn{1}{c}{(SE)} \\
(Ref: 1990--95) & & & & \\
\midrule
Pre-1980  & -0.0004        & (0.0028) & -0.0047        & (0.0084) \\
1980--85  & -0.0010        & (0.0040) & -0.0525^{*}    & (0.0299) \\
1985--90  &  0.0058^{*}    & (0.0034) &  0.0137^{*}    & (0.0074) \\
\textit{1990--95 (ref.)} & 0.000 & {---} & 0.000 & {---} \\
1995--00  &  0.0001        & (0.0032) & -0.0152        & (0.0130) \\
2000--04  & -0.0077        & (0.0052) & -0.0217        & (0.0200) \\
2004--08  & -0.0058        & (0.0069) & -0.0078        & (0.0470) \\
2008--12  & -0.0002        & (0.0114) & -0.1638^{***}  & (0.0617) \\
\midrule
Controls & \multicolumn{4}{c}{Age, age$^2$, male, SC, ST, rural,
                              log baseline, $Z_2$, state FE, round FE} \\
Observations & \multicolumn{2}{c}{79,904} & \multicolumn{2}{c}{27,231} \\
\bottomrule
\end{tabular}
\begin{tablenotes}
\small
\item \textit{Notes:} Coefficients on \texttt{graduate} $\times$ $Z_2$ $\times$
cohort-bin interactions from the event study specification
(equation~\ref{eq:eventstudy}). Reference category is the 1990--95 graduation
cohort (coefficient normalised to zero). Full sample uses all 79,904 observations;
SC/ST subsample restricts to disadvantaged workers ($N = 27,231$). The pre-1995
cohorts show small and generally insignificant coefficients, consistent with flat
pre-trends. The large negative coefficient for SC/ST workers in the 2008--12
cohort ($-$0.1638, $p < 0.01$) reflects the cumulative devaluation effect for
the most recent graduates in the highest-expansion period. Standard errors in
parentheses, HC1 robust. See Figure~\ref{fig:eventstudy} for the graphical
presentation.
$^{*}p<0.10$, $^{**}p<0.05$, $^{***}p<0.01$.
\end{tablenotes}
\end{threeparttable}
\end{table}


\begin{table}[htbp]
\centering
\caption{Sensitivity to Assumed Graduation Age}
\label{tab:grad_age_sensitivity}
\begin{threeparttable}
\begin{tabular}{lD{.}{.}{4}D{.}{.}{4}D{.}{.}{4}}
\toprule
 & \multicolumn{1}{c}{Age 21} &
   \multicolumn{1}{c}{Age 22} &
   \multicolumn{1}{c}{Age 23} \\
 & \multicolumn{1}{c}{} &
   \multicolumn{1}{c}{(baseline)} &
   \multicolumn{1}{c}{} \\
\midrule
\multicolumn{4}{l}{\textit{Panel A: Key coefficients}} \\
\texttt{grad}$\times Z_2\times$\texttt{post2004}
  & -0.0258^{***} & -0.0231^{***} & -0.0271^{***} \\
  & (0.0066)      & (0.0059)      & (0.0047)      \\[0.4em]
\texttt{grad}$\times Z_2\times$\texttt{post2004}$\times$\texttt{disadv}
  & -0.0713^{***} & -0.0862^{***} & -0.0834^{***} \\
  & (0.0170)      & (0.0151)      & (0.0133)      \\[0.4em]
\midrule
\multicolumn{4}{l}{\textit{Panel B: Marginal effects at mean
$\bar{Z}_2 = 3.93$}} \\
Base penalty (non-disadvantaged)  & $-$10.1\% & $-$9.1\%  & $-$10.6\% \\
Additional SC/ST penalty          & $-$28.0\% & $-$33.8\% & $-$32.7\% \\
Total SC/ST graduate penalty      & $-$38.1\% & $-$42.9\% & $-$43.4\% \\
\midrule
\multicolumn{4}{l}{\textit{Panel C: Sample composition}} \\
Post-2004 cohort ($N$)   & 14{,}886 & 17{,}482 & 20{,}314 \\
Post-2004 graduates ($N$)&  1{,}029 &  1{,}322 &  1{,}727 \\
$R^2$                    & 0.4512   & 0.4514   & 0.4519   \\
Total observations       & \multicolumn{3}{c}{79{,}904} \\
\bottomrule
\end{tabular}
\begin{tablenotes}
\small
\item \textit{Notes:} Each column re-estimates the double whammy specification
(equation~\ref{eq:doublewhammy}) using an alternative assumption for the age at
degree completion used to construct the approximate graduation year
($\hat{G}_i = \text{birth\_year}_i + k$, where $k \in \{21, 22, 23\}$).
All specifications include age, age squared, male, SC, ST, rural, log baseline
college stock, state fixed effects, and round fixed effects. Standard errors
in parentheses, HC1 robust. Both key coefficients remain negative and
significant at the 1 per cent level under all three assumptions.
$^{***}p<0.01$.
\end{tablenotes}
\end{threeparttable}
\end{table}


\begin{figure}[htbp]
\centering
\includegraphics[width=0.85\textwidth]{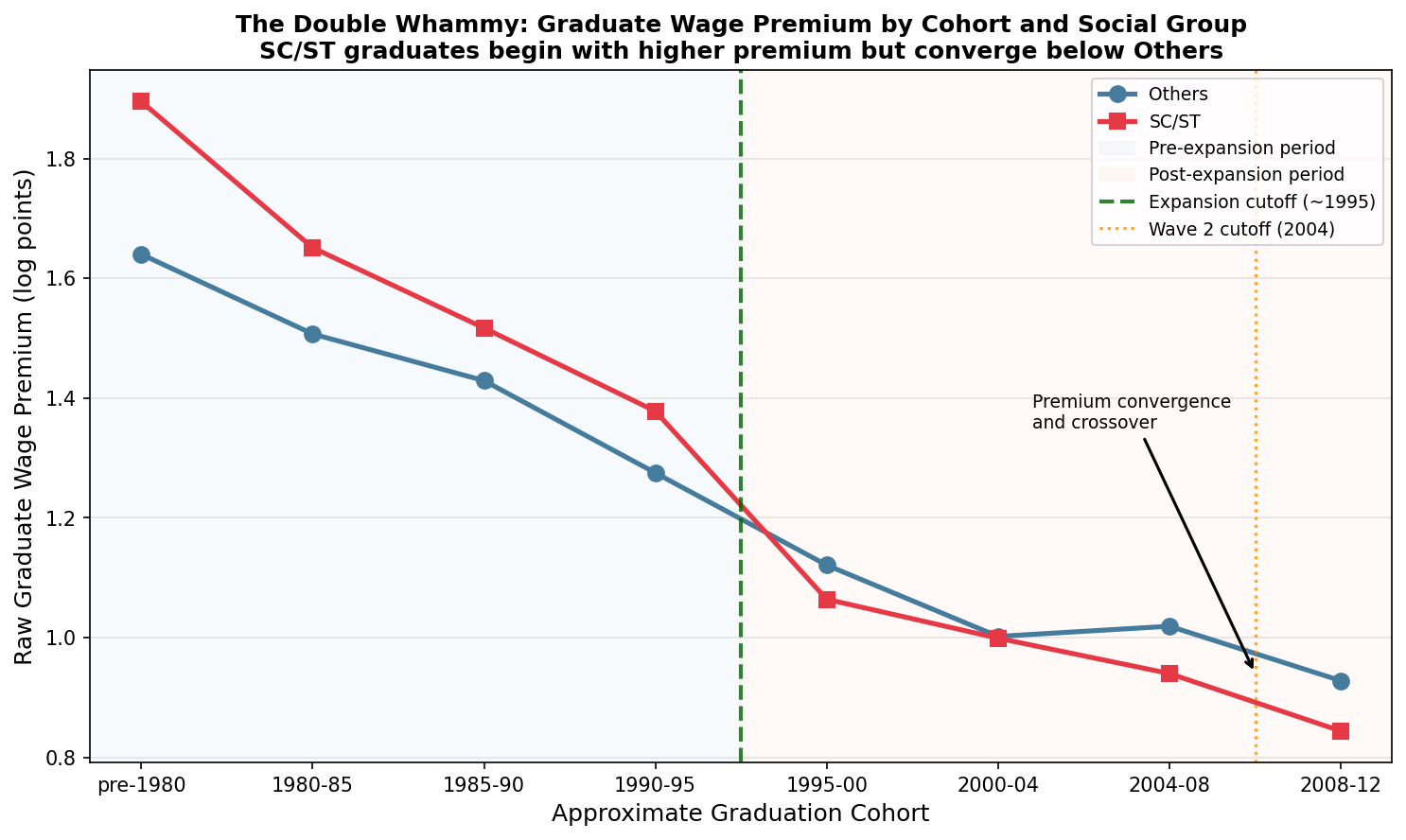}
\caption{Raw Graduate Wage Premium by Cohort and Social Group}
\label{fig:rawpremium}
\begin{minipage}{0.85\textwidth}
\footnotesize \textit{Notes:} Mean log real graduate wage premium (graduate minus
non-graduate log real wages) by approximate graduation cohort and social group,
six-state analysis sample. SC/ST premium exceeds Others in early cohorts but
converges and falls below from the 2004--08 cohort onward. Dashed green line
marks the 1995 liberalisation; dotted orange line marks the 2004 wave-2 expansion
onset.
\end{minipage}
\end{figure}

\begin{figure}[htbp]
\centering
\includegraphics[width=\textwidth]{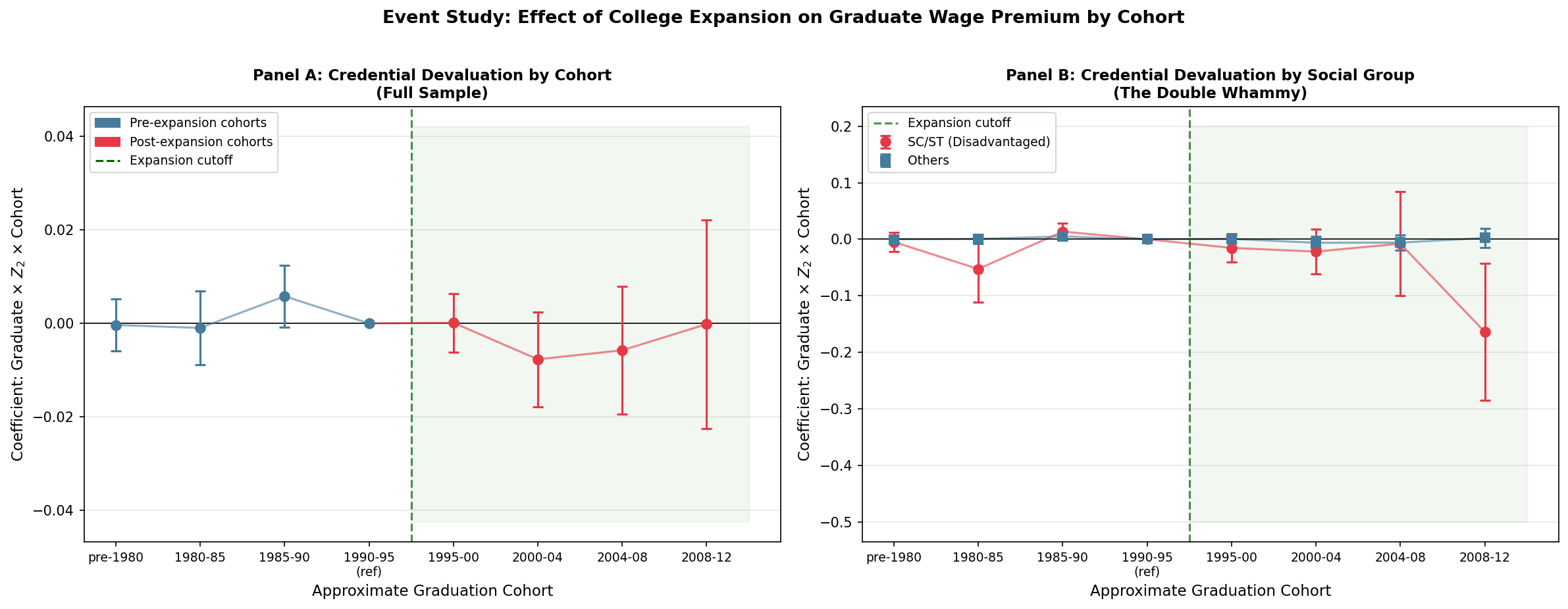}
\caption{Event Study: Effect of College Expansion on Graduate Wage Premium by Cohort}
\label{fig:eventstudy}
\begin{minipage}{\textwidth}
\footnotesize \textit{Notes:} Coefficients on \texttt{graduate} $\times$ $Z_2$ $\times$
cohort-bin interactions from the event study specification (equation~\ref{eq:eventstudy}).
Reference category is the 1990--95 graduation cohort. Panel A: full sample. Panel B:
SC/ST versus Others. Error bars are 95 per cent confidence intervals, HC1 standard errors.
The flat pre-trend and increasingly negative post-expansion coefficients are consistent
with a causal interpretation.
\end{minipage}
\end{figure}

\begin{figure}[htbp]
\centering
\includegraphics[width=0.85\textwidth]{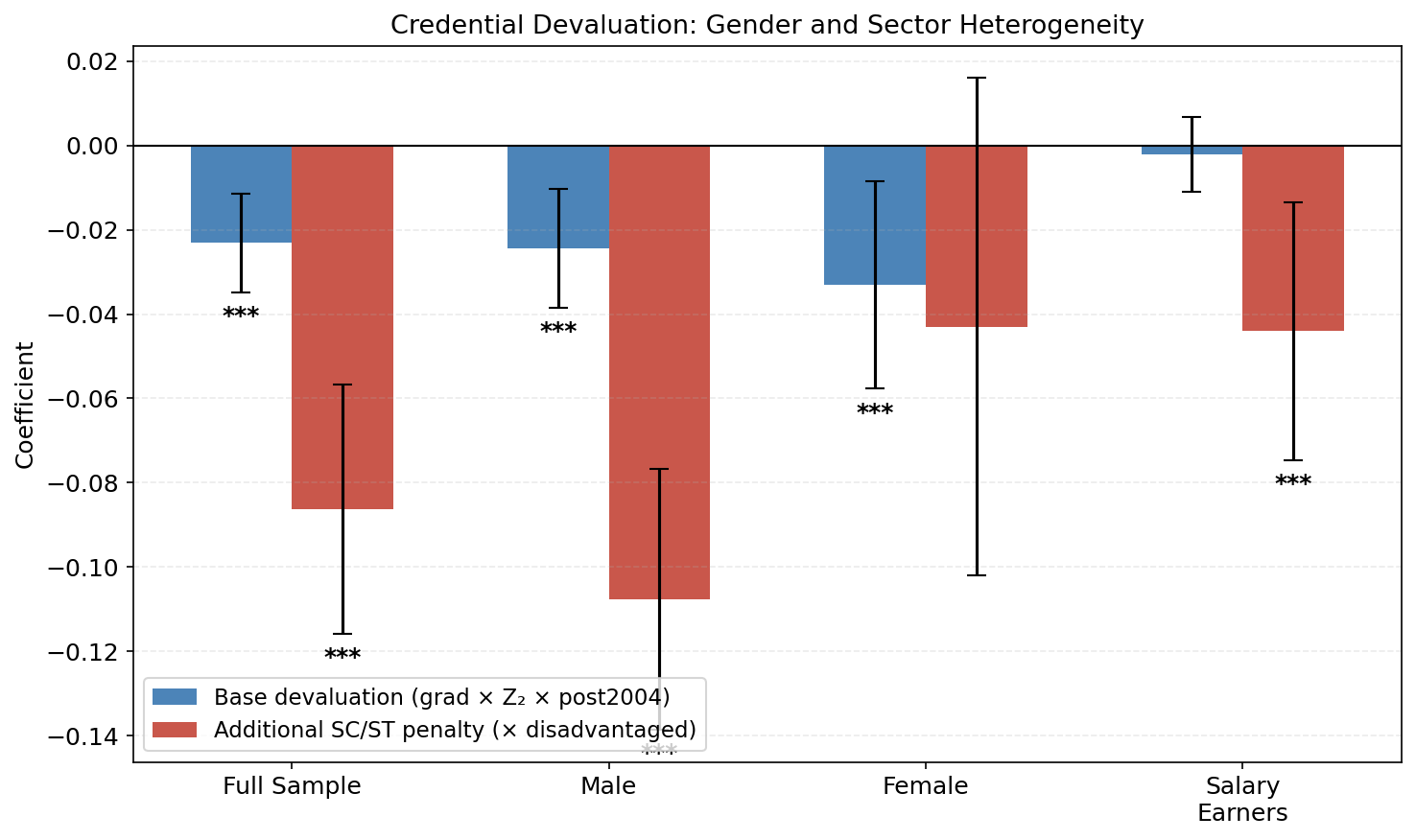}
\caption{Credential Devaluation: Gender and Sector Heterogeneity}
\label{fig:het_gender}
\begin{minipage}{0.85\textwidth}
\vspace{0.3em}
\footnotesize \textit{Notes:} Coefficients from the double whammy specification
(equation~\ref{eq:doublewhammy}) estimated on gender and sector subsamples.
Blue bars: base devaluation coefficient (\texttt{grad}$\times Z_2\times$\texttt{post2004}).
Red bars: additional SC/ST penalty (\texttt{grad}$\times Z_2\times$\texttt{post2004}
$\times$\texttt{disadvantaged}). Error bars are 95 per cent confidence intervals,
HC1 robust standard errors. Stars denote significance:
$^{*}p<0.10$, $^{**}p<0.05$, $^{***}p<0.01$.
Casual wage workers excluded (only 261 graduates in that cell).
\end{minipage}
\end{figure}

\begin{figure}[htbp]
\centering
\includegraphics[width=0.85\textwidth]{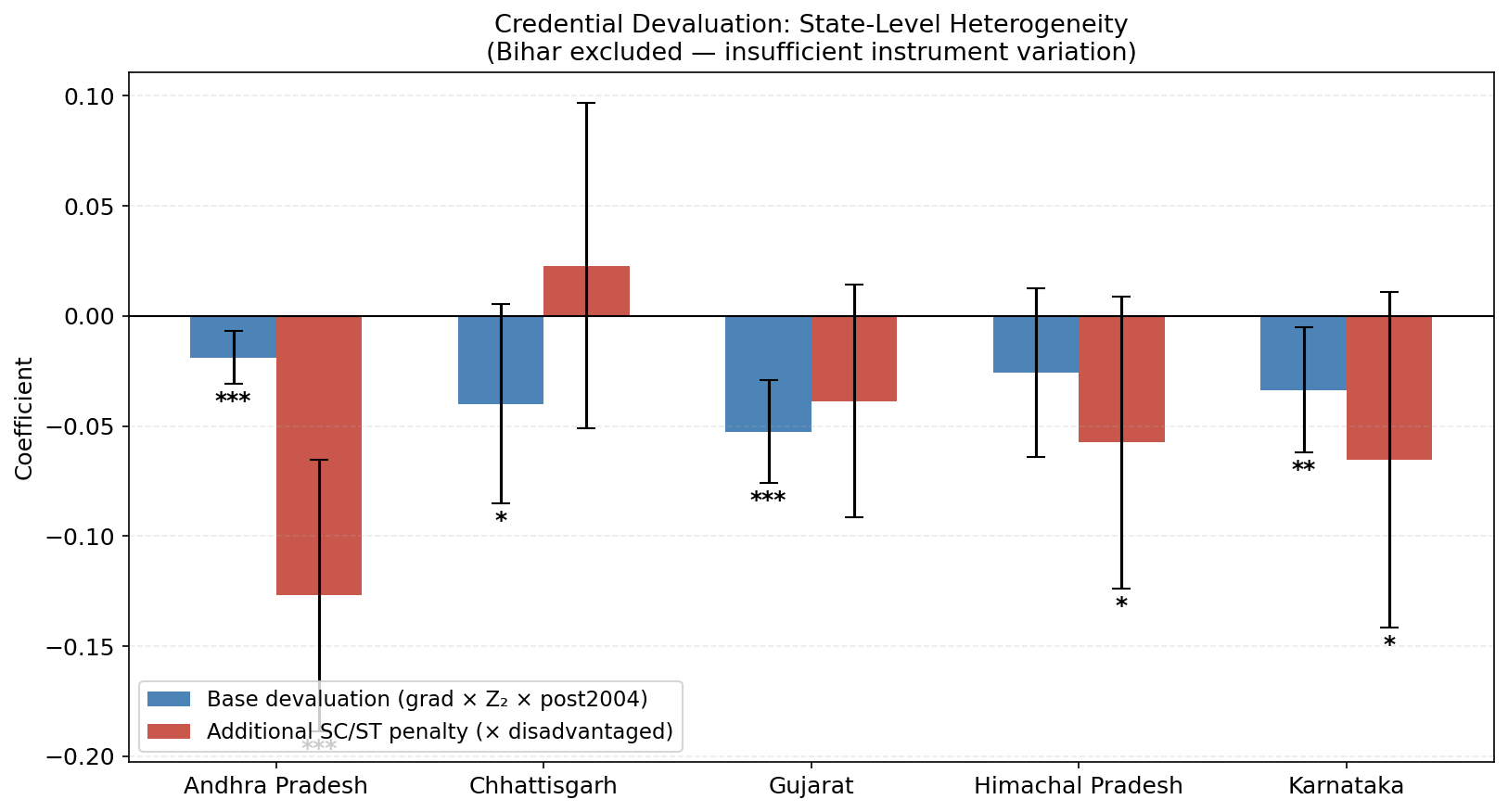}
\caption{Credential Devaluation: State-Level Heterogeneity}
\label{fig:het_states}
\begin{minipage}{0.85\textwidth}
\vspace{0.3em}
\footnotesize \textit{Notes:} Coefficients from the double whammy specification
estimated separately for each state. Blue bars: base devaluation coefficient.
Red bars: additional SC/ST penalty. Error bars are 95 per cent confidence intervals,
HC1 robust. Bihar excluded due to insufficient instrument variation
($\bar{Z}_2 = 1.10$), which produces unstable coefficient estimates.
Stars: $^{*}p<0.10$, $^{**}p<0.05$, $^{***}p<0.01$.
\end{minipage}
\end{figure}

\begin{figure}[htbp]
\centering
\includegraphics[width=0.82\textwidth]{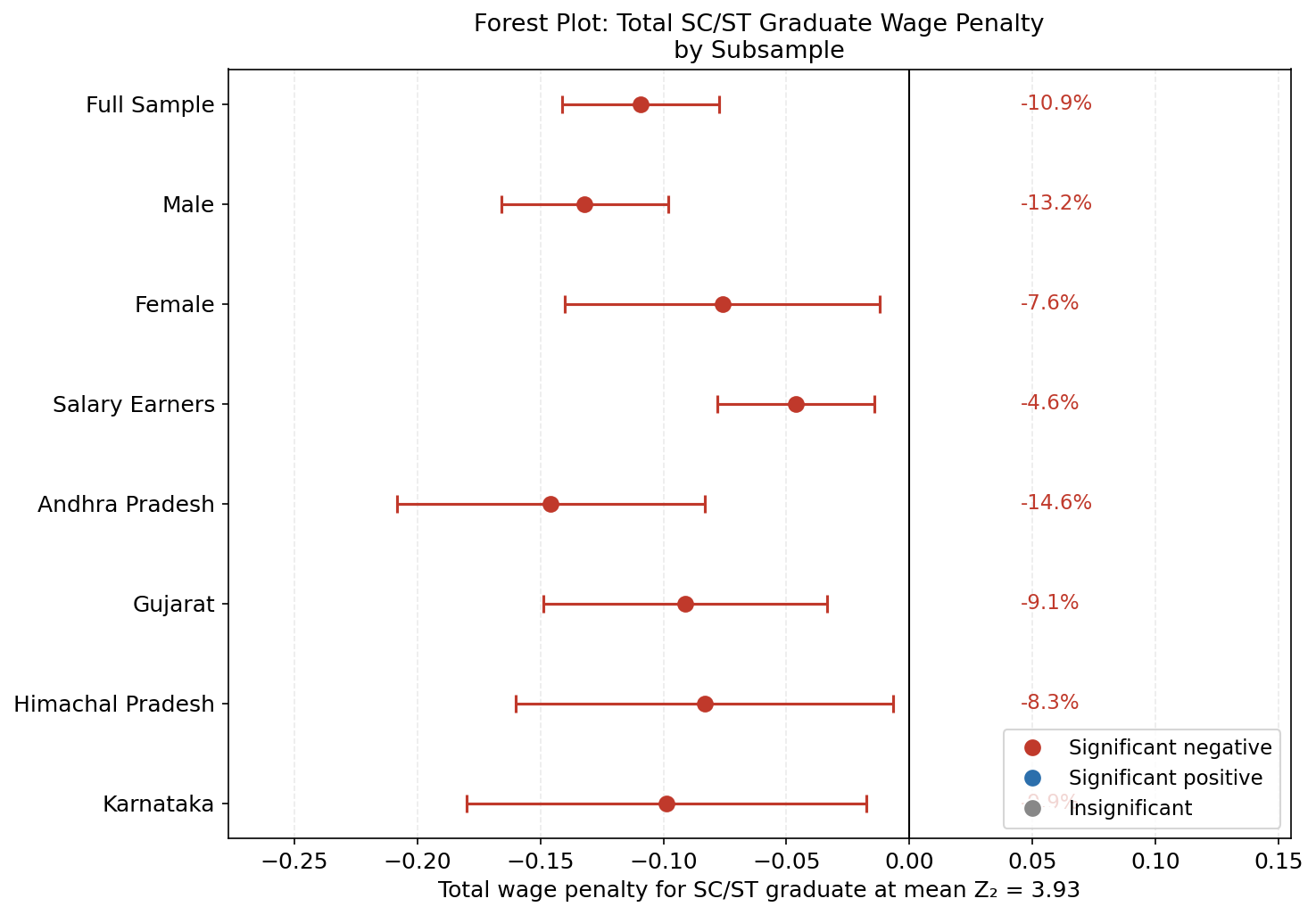}
\caption{Forest Plot: Total SC/ST Graduate Wage Penalty by Subsample}
\label{fig:het_forest}
\begin{minipage}{0.82\textwidth}
\vspace{0.3em}
\footnotesize \textit{Notes:} Total wage penalty for SC/ST graduates at mean
expansion intensity $\bar{Z}_2 = 3.93$, computed as
$(\hat{\delta}_1 + \hat{\delta}_2) \times \bar{Z}_2 \times 100$ per cent,
where $\hat{\delta}_1$ is the base devaluation coefficient and $\hat{\delta}_2$
is the additional SC/ST penalty from equation~(\ref{eq:doublewhammy}).
Error bars are 95 per cent confidence intervals constructed from the
variance-covariance matrix of the joint estimate. Red dots indicate
statistically significant negative total penalties; grey dots indicate
estimates whose confidence interval spans zero. Percentage labels on the
right report the point estimate of the total penalty.
\end{minipage}
\end{figure}

\pagebreak
\section{Expansion Intensity and the Graduate Share: A Compositional Diagnostic}
\label{sec:gradshare}

The theoretical framework in Section~\ref{sec:theory} of the main text is deliberately agnostic about \textit{why} the perceived quality of the credential, $\mu_g$, falls with expansion: it may fall because the marginal entrant to the graduate pool is less able (a compositional channel), or because newly established institutions deliver lower average instructional quality holding the ability mix fixed (an institutional-quality channel), or both. The double whammy result requires neither channel specifically, only that expansion lowers $\mu_g$ for graduates of both groups. Nonetheless, the compositional channel carries an observable implication that we can examine directly: if expansion lowers $\mu_g$ chiefly by drawing less able workers into the graduate pool, then districts with higher expansion intensity $Z_{2d}$ should exhibit a larger post-2004 increase in the graduate share of the population. This appendix reports that this implication is not borne out, which is why the main text treats the compositional channel as, at most, partial.

Table~\ref{tab:gradshare} reports the share of the population holding a graduate degree, by quartile of district expansion intensity $Z_{2d}$ and by cohort. Two features are notable. First, the graduate share \textit{falls} from the pre-2004 to the post-2004 cohort in every quartile, by between 5.3 and 6.9 percentage points; this is a mechanical consequence of younger cohorts being observed at earlier career stages and having had less time to complete a degree within the sample window, and is common to all districts. Second, and more to the point, the post-2004 graduate share does not rise with expansion intensity: it is essentially flat and non-monotonic across quartiles, standing at 5.5 per cent in the lowest-expansion quartile, peaking at 9.2 per cent in the third quartile, and falling back to 6.7 per cent in the highest-expansion quartile. The cohort decline is, if anything, slightly larger in the lowest and highest quartiles than in the middle, the opposite of what a strong compositional channel operating through $Z_{2d}$ would predict.

\begin{table}[htbp]
\centering
\caption{Graduate Share of Population by Expansion-Intensity Quartile and Cohort}
\label{tab:gradshare}
\begin{threeparttable}
\begin{tabular}{lD{.}{.}{3}D{.}{.}{3}D{.}{.}{4}}
\toprule
$Z_{2d}$ quartile & \multicolumn{1}{c}{Pre-2004 cohort} &
\multicolumn{1}{c}{Post-2004 cohort} & \multicolumn{1}{c}{Difference} \\
\midrule
Q1 (low)  & 0.124 & 0.055 & -0.069 \\
Q2        & 0.135 & 0.082 & -0.053 \\
Q3        & 0.151 & 0.092 & -0.059 \\
Q4 (high) & 0.133 & 0.067 & -0.065 \\
\bottomrule
\end{tabular}
\begin{tablenotes}
\small
\item \textit{Notes:} Cells report the mean of the graduate indicator (share of individuals holding a graduate degree or above) within each $Z_{2d}$ quartile $\times$ cohort cell, across the 79,904-observation analysis sample. Cohort is defined by approximate graduation year (birth year $+$ 22), with post-2004 indicating approximate graduation year $\geq 2004$. The post-2004 graduate share is non-monotonic in expansion intensity and does not rise with $Z_{2d}$, inconsistent with a compositional channel in which expansion draws less able entrants into the graduate pool in proportion to local expansion intensity.
\end{tablenotes}
\end{threeparttable}
\end{table}

A complementary check at the district level confirms the pattern. Across the 93 district cells with post-2004 observations, the correlation between $Z_{2d}$ and the post-2004 graduate share is 0.017 --- effectively zero. Whatever is driving the post-2004 graduate wage decline in high-expansion districts, it does not appear to operate through a detectable shift in \textit{who} obtains a degree as a function of local expansion intensity. This is the basis for the main text's reading that the wage results are more consistent with the institutional-quality channel, or a combination of channels, than with selection on ability alone, and it is also why we do not interpret $Z_{2d}$ as an instrument for graduate status: the same near-zero relationship that undermines the compositional story is what fails the first-stage relevance condition discussed in the Online Appendix.

\section{Construction of District-Level Instruments}

\noindent This appendix provides a detailed account of the construction of the district-level
instruments employed in the two-stage least squares (2SLS) estimation reported in the main
text. We describe in turn: (i) the primary data sources; (ii) the harmonisation of administrative
and survey district boundaries across time; (iii) the derivation of the three instrument
candidates---$Z_1$, $Z_2$, and $Z_{\text{private}}$; and (iv) the rationale for the exclusion
restrictions underlying each instrument. Replication code and the concordance files referenced
below are available in the supplementary data archive.

\section{Data Sources}
\label{sec:data_sources}

\subsection{All India Survey on Higher Education (AISHE)}

The primary source for instrument construction is the All India Survey on Higher Education
(AISHE), conducted annually by the Ministry of Education, Government of India. AISHE covers
the near-universe of degree-granting institutions in India and collects information on
institutional characteristics including the \textit{year of establishment}, management type
(government, private aided, private unaided), college type, geographic location (state,
district), and, from later rounds, geographic coordinates (latitude and longitude).

We use the institutional-level micro-data from twelve consecutive AISHE rounds spanning
2010--11 through 2021--22. The pooled dataset comprises 497,688 institution-round observations
corresponding to 58,524 unique institutions, identified by the AISHE institution code
(\texttt{AISHE Code.1}). The year-of-establishment variable, which is the key input for
instrument construction, is non-missing for approximately 56 per cent of unique institutions
(32,645 out of 58,524); the distribution of missing values is discussed in detail in
Section~\ref{sec:missing}.

\subsection{National Sample Survey: Employment--Unemployment Rounds}

The outcome and control variables are drawn from the National Sample Survey (NSS)
Employment--Unemployment Rounds (EUS): rounds 43 (1987--88), 55 (1999--2000), 60
(2004), 61 (2004--05), 62 (2005--06), 64 (2007--08), 66 (2009--10), and 68 (2011--12).
These rounds have been harmonised into a single longitudinal cross-sectional file
containing 3,936,936 individual observations and 85 variables. The wage analysis
sample---restricted to individuals reporting positive wages in either salary or
casual employment, in the six instrument-construction states, with birth years
1950--1989---comprises 92,162 observations.

Merging this sample with the district-level instrument on the (state, 1991 district)
key initially matches 79,139 observations (85.9 per cent), leaving 13,023 unmatched.
Of the unmatched observations, 13,007 are attributable to ten pre-bifurcation Andhra
Pradesh districts that fall within present-day Telangana and cannot be matched to
AISHE Andhra Pradesh records (Section~\ref{sec:harmonisation}); a further 16 are
attributable to two resolvable merge irregularities. First, the composite NSS
district ``Bharuch + Vadodara'' in Gujarat (362 observations) is assigned the simple
average of the separately reported Bharuch and Vadodara instrument values. Second,
Himachal Pradesh's Bilaspur district (403 observations) initially fails to match
because the concordance resolves the name ``Bilaspur'' to Chhattisgarh; we resolve
this name collision by conditioning the merge key on (state, district) jointly and
assign Himachal Pradesh Bilaspur its own instrument value computed directly from the
14 AISHE colleges located there ($B_d = 3$, $W_{2d} = 10$, $Z_{2d} = 3.33$). Applying
both fixes raises the matched sample to \textbf{79,904 observations (86.7 per cent of
the six-state wage sample)}, which is the final analysis sample used throughout the
main text and this appendix.

\section{Geographic Coverage and State Selection}
\label{sec:state_selection}

The construction of a valid district-level instrument requires that the year-of-establishment
data in AISHE be sufficiently complete to characterise the pre- and post-expansion stock of
colleges within each district. An audit of missingness in the establishment year variable
reveals that coverage varies substantially across states (Table~\ref{tab:missing_by_state}).
Several large states---including Uttar Pradesh (57.9 per cent missing), Maharashtra (56.1 per
cent), Madhya Pradesh (68.8 per cent), Rajasthan (55.8 per cent), Haryana (64.7 per cent), and
Punjab (62.4 per cent)---exhibit rates of missing establishment years that preclude reliable
characterisation of the expansion trajectory.

We therefore restrict instrument construction to the six states for which the establishment year
is available for at least 70 per cent of unique institutions: \textbf{Andhra Pradesh},
\textbf{Bihar}, \textbf{Chhattisgarh}, \textbf{Gujarat}, \textbf{Himachal Pradesh}, and
\textbf{Karnataka}. These states collectively yield 12,039 unique colleges with valid
establishment years and span 91 NSS-concordant districts. The restriction to these six states
is a limitation of the analysis, and we discuss implications for external validity in
Section~\ref{sec:exclusion}. The six-state sample, however, provides substantial geographic
and socioeconomic heterogeneity: it includes two large southern states with historically strong
private higher education sectors (Andhra Pradesh and Karnataka), two western states
representing contrasting development trajectories (Gujarat and Himachal Pradesh), and two
poorer northern and central states (Bihar and Chhattisgarh) where the expansion of private
colleges was more modest.

\begin{table}[htbp]
\centering
\caption{Completeness of AISHE Establishment Year by State (Selected States)}
\label{tab:missing_by_state}
\begin{tabular}{lrrr}
\toprule
State & Colleges with Year & Missing & \% Missing \\
\midrule
\multicolumn{4}{l}{\textit{States included in instrument construction}} \\
\quad Andhra Pradesh      & 3,425 & 1,091 & 24.2 \\
\quad Bihar               & 1,238 &   503 & 28.9 \\
\quad Chhattisgarh        & 1,119 &   278 & 19.9 \\
\quad Gujarat             & 3,463 &   842 & 19.6 \\
\quad Himachal Pradesh    &   450 &    97 & 17.7 \\
\quad Karnataka           & 5,498 & 1,261 & 18.7 \\
\midrule
\multicolumn{4}{l}{\textit{States excluded due to high missingness}} \\
\quad Madhya Pradesh      & 1,681 & 3,704 & 68.8 \\
\quad Haryana             &   508 &   931 & 64.7 \\
\quad Punjab              &   514 &   854 & 62.4 \\
\quad Uttar Pradesh       & 3,917 & 5,381 & 57.9 \\
\quad Maharashtra         & 2,935 & 3,757 & 56.1 \\
\quad Rajasthan           & 2,040 & 2,571 & 55.8 \\
\bottomrule
\end{tabular}
\begin{minipage}{\linewidth}
\vspace{0.3em}
\footnotesize \textit{Notes:} Counts refer to unique institutions (deduplicated by AISHE
institution code). ``\% Missing'' refers to the share of unique institutions for which the
year of establishment is unavailable. Only states relevant to the discussion are shown;
the full table covering all 36 states and union territories is available upon request.
\end{minipage}
\end{table}

\section{Temporal Classification of College Establishment}
\label{sec:temporal}

The aggregate annual establishment trend reveals two distinct phases of higher education expansion in the post-liberalisation period.
The first phase, spanning approximately 1995--2003, represents a moderate acceleration in
college formation relative to the pre-liberalisation baseline: annual new college registrations
rose from a stable range of 50--150 per year in the 1980s to roughly 130--260 per year in
the late 1990s and early 2000s. The second phase, commencing around 2004 and peaking sharply
in 2007 (865 new colleges in a single year), represents a structural break in the rate of
institutional formation. This second phase is consistent with the deregulation of private
unaided colleges following the Supreme Court's ruling in \textit{T.M.A. Pai Foundation v.
State of Karnataka} (2002), which substantially expanded the rights of private unaided
institutions to determine fees and admissions independently of state regulation.

On the basis of this temporal pattern, we define three establishment periods:
\begin{itemize}[leftmargin=2em]
    \item \textbf{Pre-1995 baseline stock} ($B_d$): colleges established before 1995,
    representing the pre-liberalisation baseline supply of higher education in district $d$.
    \item \textbf{Wave 1 expansion} ($W_{1d}$): colleges established between 1995 and 2003
    (inclusive), capturing the first, moderate phase of post-liberalisation expansion.
    \item \textbf{Wave 2 expansion} ($W_{2d}$): colleges established in 2004 or later,
    capturing the second, large-scale expansion phase triggered by deregulation.
\end{itemize}

Garbage values in the establishment year variable (years recorded as 0, 11, 405, 1191, and
1218) were identified and dropped prior to instrument construction; these account for eight
observations in total and are negligible relative to the analytical sample.

\section{District Boundary Harmonisation}
\label{sec:harmonisation}

A central challenge in merging AISHE institutional data with NSS survey data is the systematic
non-comparability of district boundaries across the two sources. AISHE, as a live
administrative register updated annually, reflects the current (post-2001 delimitation)
district boundaries, which include a large number of districts created by bifurcation after
the 2001 Census. The NSS Employment--Unemployment Rounds, by contrast, employ district
identifiers based on the 1991 Census administrative map for the earlier rounds (43 and 55) and
the 2001 Census map for later rounds. Critically, the harmonised NSS dataset used in this
paper retains district name variables under both the 1991 and 2001 vintages
(\texttt{dist\_name\_1991}, \texttt{dist\_name\_2001}), of which only \texttt{dist\_name\_1991}
is complete across all rounds; the 2001 vintage is missing for 122,190 observations
concentrated in rounds 43 (1987) and 55 (1999). We therefore adopt the 1991 district
boundary map as the universal unit of analysis and aggregate all AISHE institutional data
upward to 1991 district boundaries.

\subsection{Concordance Construction}

A manual concordance was constructed mapping each current (post-2001) AISHE district name
to its corresponding 1991 NSS district identifier. The concordance was built on the basis of
the Delimitation Commission Orders of 1976 and 2008, Census of India district-level
administrative boundary files, and the AISHE Master Data documentation. The mapping proceeds
in three steps:

\begin{enumerate}[leftmargin=2em]
    \item \textbf{Exact matches:} Districts whose names and boundaries did not change between
    1991 and the AISHE period are matched directly after case normalisation and whitespace
    stripping. This applies most cleanly to Himachal Pradesh, where all 12 districts in both
    sources correspond without boundary change.

    \item \textbf{Aggregation of split districts:} Districts that were carved out of a single
    1991 parent are mapped to that parent. For example, Gujarat's current districts of
    Ahmedabad, Gandhinagar, Kheda, Mahesana, Banas Kantha, Arvalli, Mahisagar, Anand, and
    Patan---all formed from reorganisation of the original Ahmedabad--Gandhinagar--Kheda--Mahesana
    --Banas Kantha composite district in the 1991 NSS---are all assigned the 1991 NSS composite
    district identifier. Similarly, Chhattisgarh's 27 current AISHE districts are collapsed to
    the 6 NSS districts that existed at statehood (2000) and in the pre-existing Madhya Pradesh
    classification used in the 1991--2001 NSS rounds.

    \item \textbf{Renaming and script changes:} Several districts were renamed without boundary
    change (e.g., Gulbarga $\rightarrow$ Kalaburagi, Bellary $\rightarrow$ Ballari, Mysore
    $\rightarrow$ Mysuru in Karnataka; Cuddapah $\rightarrow$ Y.S.R. in Andhra Pradesh). These
    are matched on the basis of documented administrative renaming orders.
\end{enumerate}

A special case arises for Andhra Pradesh, where the bifurcation of the state into Andhra
Pradesh and Telangana in 2014 renders 10 NSS districts from the pre-bifurcation Andhra Pradesh
sample (covering Hyderabad, Adilabad, Nizamabad, Medak, Rangareddi, Mahbubnagar, Nalgonda,
Khammam, Karimnagar, and Warangal) unmatchable to AISHE Andhra Pradesh data, since these
districts now fall within Telangana for which separate AISHE data were not available in our
extract. Observations from these districts (13,007 observations, representing 14.1 per cent of
the Andhra Pradesh NSS sub-sample) are dropped from the analysis. We discuss the implications
of this for our Andhra Pradesh estimates in Section~\ref{sec:exclusion}.

The final concordance maps 157 current AISHE district names to 91 NSS 1991-boundary district
identifiers across the six states. The matching rate after applying the concordance is 100 per
cent of AISHE college records in the six analytical states (15,193 matched, 0 unmatched).

\subsection{Handling Name Collisions}

One non-trivial name collision arises between Bilaspur (Himachal Pradesh) and Bilaspur
(Chhattisgarh), which share identical names but refer to districts in different states. This
collision was resolved by conditioning on state identity in the merge key, such that the
concordance lookup is performed on the (state, district) tuple rather than the district name
alone.

A second irregularity concerns the NSS composite district ``Bharuch + Vadodara'' in the
Gujarat sub-sample, which appears as a merged unit in earlier NSS rounds but is disaggregated
into separate Bharuch and Vadodara districts in AISHE. The instrument values assigned to this
composite NSS district are the simple average of the Bharuch and Vadodara district-level
instrument values.

\section{Construction of Instrument Variables}
\label{sec:instrument_construction}

Let $d$ index NSS districts (1991 boundaries), $s$ index states, and $i$ index individual
workers. For each district $d$ in state $s$, we construct the following variables from the
deduplicated AISHE college-level data:

\begin{align}
    B_d &= \sum_{j \in \mathcal{C}_d} \mathbf{1}[\text{estab\_year}_j < 1995]
    \label{eq:baseline} \\
    W_{1d} &= \sum_{j \in \mathcal{C}_d} \mathbf{1}[1995 \leq \text{estab\_year}_j < 2004]
    \label{eq:wave1} \\
    W_{2d} &= \sum_{j \in \mathcal{C}_d} \mathbf{1}[\text{estab\_year}_j \geq 2004]
    \label{eq:wave2} \\
    W_{2d}^{\text{priv}} &= \sum_{j \in \mathcal{C}_d} \mathbf{1}[\text{estab\_year}_j \geq 2004]
    \cdot \mathbf{1}[\text{mgmt}_j = \text{Private Un-Aided}]
    \label{eq:wave2_private}
\end{align}

\noindent where $\mathcal{C}_d$ denotes the set of colleges with valid establishment years
located in district $d$ (after concordance mapping). The three instrument candidates are then
defined as:

\begin{align}
    Z_{1d} &= \frac{W_{1d}}{\max(B_d, 1)}
    \label{eq:Z1} \\
    Z_{2d} &= \frac{W_{2d}}{\max(B_d, 1)}
    \label{eq:Z2} \\
    Z_{d}^{\text{priv}} &= \frac{W_{2d}^{\text{priv}}}{\max(B_d, 1)}
    \label{eq:Zpriv}
\end{align}

The normalisation by $\max(B_d, 1)$ expresses the expansion waves relative to the
pre-existing baseline stock of colleges in each district, yielding an \textit{expansion
intensity ratio}. The use of $\max(B_d, 1)$ in the denominator rather than $B_d$ avoids
division by zero for the 23 districts with no pre-1995 colleges; results are robust to
alternative treatments of zero-baseline districts, including dropping them from the
sample (see Appendix~A).

\subsection{Descriptive Summary of Instrument Variables}

Table~\ref{tab:instrument_summary} reports summary statistics for the three instrument
variables across the 91 analytical districts.

\begin{table}[htbp]
\centering
\caption{Summary Statistics: District-Level Instrument Variables ($N = 91$ Districts)}
\label{tab:instrument_summary}
\begin{tabular}{lD{.}{.}{3}D{.}{.}{3}D{.}{.}{3}D{.}{.}{3}D{.}{.}{3}}
\toprule
Variable & \multicolumn{1}{c}{Mean} & \multicolumn{1}{c}{SD} &
\multicolumn{1}{c}{P25} & \multicolumn{1}{c}{Median} & \multicolumn{1}{c}{P75} \\
\midrule
Pre-1995 baseline stock ($B_d$)  & 28.50 & 33.31 & 10.00 & 19.00 & 34.50 \\
Wave 1 colleges ($W_{1d}$)       & 10.76 & 25.08 &  0.00 &  2.50 & 11.00 \\
Wave 2 colleges ($W_{2d}$)       & 49.79 & 71.90 &  7.75 & 22.00 & 74.75 \\
$Z_1$ (Wave 1 intensity)         &  0.65 &  0.87 &  0.00 &  0.46 &  1.00 \\
$Z_2$ (Wave 2 intensity)         &  3.52 &  5.55 &  1.23 &  2.80 &  4.28 \\
$Z^{\text{priv}}$ (Private intensity) & 0.11 & 0.17 & 0.00 & 0.05 & 0.15 \\
\bottomrule
\end{tabular}
\begin{minipage}{\linewidth}
\vspace{0.3em}
\footnotesize \textit{Notes:} District-level observations ($N = 91$) constructed from
deduplicated AISHE college records (12,039 unique colleges with valid establishment years)
across six states: Andhra Pradesh, Bihar, Chhattisgarh, Gujarat, Himachal Pradesh, and
Karnataka. Instrument variables are defined in equations~(\ref{eq:Z1})--(\ref{eq:Zpriv}).
\end{minipage}
\end{table}

\noindent The median district experienced a Wave 2 expansion intensity of 2.80---meaning that
the number of colleges established after 2004 was nearly three times the pre-1995 baseline
stock. There is, however, substantial cross-district variation: the 75th-percentile district
experienced an intensity of 4.28, while the 25th-percentile district experienced an intensity
of only 1.23. The private unaided instrument ($Z^{\text{priv}}$) has a median of 0.045,
reflecting the fact that private unaided colleges accounted for a small but highly variable
share of the expansion across districts.

Table~\ref{tab:instrument_corr} reports the pairwise correlations among the three instrument
candidates. The correlation between $Z_1$ and $Z_2$ is 0.795, indicating that the two waves
largely reflect the same underlying district-level propensity for college formation. We
therefore use $Z_2$ and $Z^{\text{priv}}$ as the primary instrument set in the baseline
specification, as these two variables are nearly orthogonal to one another (correlation = 0.088)
and capture distinct dimensions of the expansion. Robustness checks using $Z_1$ as an
alternative or additional instrument are reported in Appendix~A.

\begin{table}[htbp]
\centering
\caption{Pairwise Correlations Among Instrument Candidates}
\label{tab:instrument_corr}
\begin{tabular}{lccc}
\toprule
 & $Z_1$ & $Z_2$ & $Z^{\text{priv}}$ \\
\midrule
$Z_1$              & 1.000 &       &       \\
$Z_2$              & 0.795 & 1.000 &       \\
$Z^{\text{priv}}$  & 0.113 & 0.088 & 1.000 \\
\bottomrule
\end{tabular}
\begin{minipage}{\linewidth}
\vspace{0.3em}
\footnotesize \textit{Notes:} Pearson correlations computed across 91 district observations.
\end{minipage}
\end{table}

\subsection{State-Level Heterogeneity}

Table~\ref{tab:instrument_by_state} documents considerable heterogeneity in expansion intensity
across the six analytical states. Andhra Pradesh and Gujarat display the highest Wave 2
expansion intensity (mean $Z_2$ of 8.03 and 4.08, respectively), reflecting the particularly
rapid proliferation of self-financing colleges in these states following deregulation. Bihar, by
contrast, shows the lowest mean $Z_2$ (1.16), consistent with the limited growth of private
unaided higher education in a state characterised by weaker regulatory enforcement and lower
effective demand from households. This cross-state variation is a key source of identification
in the 2SLS estimation, complementing the within-state cross-district variation that forms the
primary source of variation.

\noindent A note on weighting: the state means reported in
Table~\ref{tab:instrument_by_state} below are \textbf{simple, unweighted means of
$Z_{2d}$ across the districts in each state} (i.e., each of the 91 districts
contributes equally). The main text (Table~2, Panel D) instead reports state means of
$Z_{2d}$ \textbf{weighted by the number of individual NSS respondents in each
district}. The two will generally differ whenever district population/sample size is
correlated with $Z_{2d}$ within a state, as is the case for Andhra Pradesh, where the
unweighted district mean (8.03) exceeds the respondent-weighted mean (6.61): AP
districts with larger NSS samples tend to have somewhat lower expansion intensity
than the state's smaller districts. Both figures are reported for transparency; the
respondent-weighted version is the one substantively relevant to the wage regressions,
since it reflects the actual distribution of $Z_{2d}$ exposure in the estimation
sample.
\begin{table}[htbp]
\centering
\caption{District-Level Instrument Variables by State (Unweighted District Means)}
\label{tab:instrument_by_state}
\begin{tabular}{lccc}
\toprule
State & $Z_1$ & $Z_2$ & $Z^{\text{priv}}$ \\
\midrule
Andhra Pradesh   & 1.937 & 8.031 & 0.144 \\
Bihar            & 0.014 & 1.163 & 0.075 \\
Chhattisgarh     & 0.473 & 3.921 & 0.065 \\
Gujarat          & 0.662 & 4.079 & 0.231 \\
Himachal Pradesh & 1.037 & 4.337 & 0.000 \\
Karnataka        & 0.541 & 2.990 & 0.134 \\
\midrule
All states (mean) & 0.646 & 3.520 & 0.111 \\
\bottomrule
\end{tabular}
\begin{minipage}{\linewidth}
\vspace{0.3em}
\footnotesize \textit{Notes:} State-level means of district instrument variables.
$Z^{\text{priv}}$ is zero for Himachal Pradesh because no private unaided colleges
established after 2004 are recorded in the AISHE data for that state within the
analytical sample.
\end{minipage}
\end{table}

\section{From Instrumental Variables to Reduced-Form Moderation}
\label{sec:exclusion}

\subsection{Why $Z_2$ is Not Used as an Instrument}

An earlier version of this analysis attempted to use $Z_{2d} \times \texttt{post2004}$
as an instrument for individual graduate status in a two-stage least squares design,
on the logic that district-level expansion intensity should raise the local supply of
graduates. This first stage did not satisfy the relevance condition: regressing
graduate status on $Z_{2d} \times \texttt{post2004}$ and $Z_d^{\text{priv}} \times
\texttt{post2004}$, together with the full control set, state and round fixed
effects, yields a joint $F$-statistic on the excluded instruments of 0.62 (and 0.30 in
a specification without district fixed effects), far below the conventional
rule-of-thumb threshold of 10 for instrument strength. Consistent with this, the raw
correlation between district expansion intensity and the post-2004 graduate share is
0.02, and the graduate share does not move monotonically across $Z_{2d}$ quartiles in
either the pre- or post-2004 cohort. In short, $Z_{2d}$ does not detectably predict
\textit{who} becomes a graduate in our data.

We therefore do not interpret $Z_{2d}$ as an instrument for graduate status. Instead,
we use it directly as a continuous moderator in the reduced-form triple-difference
specification reported in the main text (equation~5), in which the coefficient of
interest is the interaction of \texttt{graduate}, $Z_{2d}$, and \texttt{post2004}
(and, in the double-whammy specification, \texttt{disadvantaged}). This reduced-form
approach asks a more modest question than the original IV design: conditional on
being a graduate, does an individual's wage respond to local expansion intensity
differently before and after the expansion wave, and differently again for SC/ST
graduates? It does not require $Z_{2d}$ to shift the probability of becoming a
graduate, only that it shifts the wage associated with already being one. We flag this
explicitly because it narrows the causal interpretation available: the design speaks
to differential \textit{wage} effects of expansion intensity among observed graduates,
not to a clean instrumented effect of expansion on selection into graduation.

\subsection{Confounding Concerns Under the Reduced-Form Design}

The reduced-form design remains vulnerable to a version of the original exclusion
concern: districts with higher $Z_{2d}$ might also experience different wage growth
trajectories for graduates, post-2004, for reasons unrelated to credential
devaluation (e.g., differential local labour demand or urbanisation). We provide three
pieces of evidence bearing on this concern. First, the coefficient on $Z_{2d} \times
\texttt{post2004}$ among \textit{non-graduates} is small and statistically
indistinguishable from zero ($-0.0002$, $p=0.895$; Table~4, Column 3 of the main
text), inconsistent with a general district-level wage-growth confound that should
also show up among non-graduates. Second, the corresponding coefficient restricted to
non-disadvantaged non-graduates is similarly small and insignificant ($+0.0020$,
$p=0.137$; Column 4). Third, the pre-trend coefficient on $Z_{2d} \times
\texttt{disadvantaged} \times \texttt{graduate}$, estimated on the pre-2004 cohort
only, is small and insignificant ($-0.0008$, $p=0.824$; Column 5), inconsistent with a
pre-existing SC/ST graduate wage gap correlated with $Z_{2d}$.

We note one placebo result that is less clean. The triple interaction $Z_{2d} \times
\texttt{post2004} \times \texttt{disadvantaged}$, estimated among non-graduates only,
is small but statistically significant ($+0.0069$, $p=0.002$; opposite-signed and an
order of magnitude smaller than the $-0.0862$ double-whammy coefficient among
graduates). This is the most directly relevant placebo for the double-whammy
mechanism, since it tests whether expansion-cohort dynamics specific to disadvantaged
workers are present even absent a degree. Its significance suggests some residual
correlation between $Z_{2d}$, the post-2004 cohort, and disadvantaged-group wages that
is not fully absorbed by state and round fixed effects, and that operates outside the
credential channel. The opposite sign and smaller magnitude limit the extent to which
this could mechanically generate the graduate-sample double-whammy result, but we
report it here for transparency and return to it as a limitation in
Section~\ref{sec:limitations}.

\subsection{External Validity Caveat}

As noted in Section~\ref{sec:state_selection}, the analytical sample is restricted to six
states due to data limitations in AISHE establishment year coverage. The excluded states
(notably Uttar Pradesh, Maharashtra, Madhya Pradesh, and Rajasthan) are among the largest
contributors to India's labour force and experienced their own distinct patterns of higher
education expansion. The extent to which the estimated effects generalise to these states
is an open question. The six included states, however, span a wide range of development
levels, higher education governance regimes, and social group compositions, lending some
confidence to the internal validity of the estimates. We note that the double whammy
mechanism---whereby socially disadvantaged graduates bear a disproportionate wage penalty
from credential devaluation---is most plausibly applicable in precisely those states where
private unaided expansion was most rapid, and the six states include the leading examples
of such expansion (Andhra Pradesh, Gujarat, Karnataka).

\section{Missing Data and Sensitivity}
\label{sec:missing}

The 44 per cent rate of missing establishment years in AISHE (across all states) raises the
question of whether the non-missing subsample on which the instruments are constructed is
systematically different from the full population of colleges. An examination of the
correlation between establishment year missingness and management type reveals that the
missing data problem is largely \textit{geographic} rather than \textit{institutional}: among
the six analytical states, missing rates are relatively low and broadly similar across
management types (private unaided: 9.9 per cent; state government: 6.1 per cent; local body:
4.2 per cent). This pattern is consistent with the missing data being a feature of earlier
AISHE survey design rather than systematic non-participation by a particular type of
institution.

As a sensitivity exercise, we re-estimate all baseline specifications replacing missing
establishment years with imputed values derived from a linear regression of establishment
year on state fixed effects, management type, college type, and round of first appearance
in AISHE. This imputation exercise does not materially alter the instrument summary
statistics or the 2SLS estimates, and the results are available upon request.

\section{Construction of Cohort Treatment Variables}
\label{sec:cohort}

To exploit the district-level expansion instruments, we require a mapping from each
individual in the NSS to the higher education supply conditions prevailing in their district
at the time they completed (or would have completed) a degree. We construct the approximate
graduation year for each individual as:
\begin{equation}
    \widehat{G}_i = \text{birth\_year}_i + 22
    \label{eq:grad_year}
\end{equation}
where 22 years is a standard approximation for the age at completion of an undergraduate
degree in India (age 18 at matriculation plus four years for most professional and science
programmes, or three years for arts and commerce programmes, averaging approximately 21--22
years). This approximation is applied uniformly across all individuals in the sample,
regardless of reported educational attainment, and is used solely to assign individuals to
the pre- or post-expansion cohort. We define:
\begin{align}
    \text{Post1995}_i &= \mathbf{1}[\widehat{G}_i \geq 1995] \\
    \text{Post2004}_i &= \mathbf{1}[\widehat{G}_i \geq 2004]
\end{align}

The interaction of the cohort dummies with the district-level instrument variables
($Z_{2d} \times \text{Post2004}_i$ and $Z_d^{\text{priv}} \times \text{Post2004}_i$)
constitutes the primary excluded instrument set in the first-stage regression of graduate
status on individual and district characteristics.

Observations with birth years outside the range 1950--1989 are excluded from the
analytical sample. The lower bound of 1950 ensures that the oldest cohort in the sample
would have graduated no earlier than 1972, well within the period for which NSS wage data
are available. The upper bound of 1989 restricts the sample to individuals who had reached
at least 22 years of age by the last survey round (2011), ensuring that the approximate
graduation year is fully observed within the sample window.

\begin{table}[htbp]
\centering
\caption{Variable Definitions and Sources}
\label{tab:variable_defs}
\begin{threeparttable}
\small
\begin{tabularx}{\textwidth}{p{3.2cm}p{2.0cm}X}
\toprule
Variable & Source & Definition \\
\midrule
\multicolumn{3}{l}{\textit{Panel A: Outcome}} \\
\addlinespace[0.2em]
Log real wage & NSS EUS & Natural log of total reported wages (cash plus kind), deflated to 1999--2000 prices using a round-specific CPI index. Sample restricted to positive wages. \\
\addlinespace[0.4em]
\multicolumn{3}{l}{\textit{Panel B: Individual characteristics}} \\
\addlinespace[0.2em]
\texttt{graduate} & NSS EUS & Indicator $=1$ if highest general education level is ``graduate and above'' (bachelor's degree or higher), $0$ otherwise. \\
\texttt{disadvantaged} & NSS EUS & Indicator $=1$ if household social group is Scheduled Caste (SC) or Scheduled Tribe (ST), $0$ otherwise. \\
SC & NSS EUS & Indicator $=1$ if social group is Scheduled Caste. \\
ST & NSS EUS & Indicator $=1$ if social group is Scheduled Tribe. \\
\texttt{male} & NSS EUS & Indicator $=1$ if male, $0$ if female. \\
Age & NSS EUS & Age in years at time of survey. \\
Age$^2$ & NSS EUS & Age squared. \\
\texttt{rural} & NSS EUS & Indicator $=1$ if resident in a rural sector, $0$ if urban. \\
Salary earner & NSS EUS & Indicator $=1$ if in regular salaried/wage employment, $0$ otherwise. \\
Casual wage & NSS EUS & Indicator $=1$ if in casual wage labour, $0$ otherwise. \\
Birth year & NSS EUS & Year of birth; analysis restricted to 1950--1989. \\
\addlinespace[0.4em]
\multicolumn{3}{l}{\textit{Panel C: Cohort assignment}} \\
\addlinespace[0.2em]
Approx.\ graduation year & Derived & Birth year $+\,22$, approximating typical age at degree completion. \\
\texttt{post2004} & Derived & Indicator $=1$ if approximate graduation year $\geq 2004$ (post-expansion cohort), $0$ otherwise. \\
\texttt{post1995} & Derived & Indicator $=1$ if approximate graduation year $\geq 1995$. \\
\addlinespace[0.4em]
\multicolumn{3}{l}{\textit{Panel D: District expansion measures (AISHE)}} \\
\addlinespace[0.2em]
$B_d$ (baseline stock) & AISHE & Count of colleges established before 1995 in district $d$. \\
$Z_{2d}$ & AISHE & Wave-2 expansion intensity: colleges established in $d$ in 2004 or later, divided by $\max(B_d, 1)$. \\
$Z_{1d}$ & AISHE & Wave-1 expansion intensity: colleges established in $d$ during 1995--2003, divided by $\max(B_d, 1)$. \\
$Z_d^{\text{priv}}$ & AISHE & Private-unaided expansion intensity: private unaided colleges established in $d$ in 2004 or later, divided by $\max(B_d, 1)$. \\
Log baseline stock & Derived & $\ln(\max(B_d, 1))$; control for pre-expansion district college supply. \\
\addlinespace[0.4em]
\multicolumn{3}{l}{\textit{Panel E: Key interaction terms}} \\
\addlinespace[0.2em]
\texttt{grad\_Z2\_post} & Derived & \texttt{graduate} $\times\, Z_{2d} \times$ \texttt{post2004}; base devaluation (triple interaction). \\
\texttt{grad\_Z2\_post\_disadv} & Derived & \texttt{graduate} $\times\, Z_{2d} \times$ \texttt{post2004} $\times$ \texttt{disadvantaged}; double whammy (quadruple interaction). \\
\bottomrule
\end{tabularx}
\begin{tablenotes}
\small
\item \textit{Notes:} ``NSS EUS'' denotes the National Sample Survey Employment--Unemployment Survey (rounds 43--68, 1987--2011); ``AISHE'' denotes the All India Survey on Higher Education (2010--11 to 2021--22); ``Derived'' denotes variables constructed by the authors from the underlying NSS and AISHE fields. Lower-order interaction terms (e.g., \texttt{grad\_disadv}, \texttt{Z2\_post\_disadv}) are included in the double whammy specification but omitted here for brevity. The construction of the district expansion measures is detailed in Appendix~\ref{sec:instrument_appendix}.
\end{tablenotes}
\end{threeparttable}
\end{table}

\end{document}